\documentclass[amsmath,twocolumn,amssymb,nofootinbib,prd,superscriptaddress]{revtex4-1}
\pdfoutput=1

\usepackage{graphicx}
\usepackage{hyperref}
\usepackage{mathtools}
\usepackage[outdir=./]{epstopdf}
\usepackage[caption=false]{subfig}

\usepackage[usenames,dvipsnames]{color}

\begin{document}

\title{Islands of stability and recurrence times in AdS}

\author{Stephen R. Green}
\email{sgreen@perimeterinstitute.ca}
\affiliation{Perimeter Institute for Theoretical Physics, Waterloo, Ontario N2L 2Y5,
Canada}
\author{Antoine Maillard}
\email{antoine.maillard@ens.fr}
\affiliation{D\'epartement de Physique, \'Ecole Normale Sup\'erieure, 24 rue Lhomond, 
75005 Paris, France}
\author{Luis Lehner}
\email{llehner@perimeterinstitute.ca}
\affiliation{Perimeter Institute for Theoretical Physics, Waterloo, Ontario N2L 2Y5,
Canada}
\author{Steven L. Liebling}
\email{steve.liebling@liu.edu}
\affiliation{Department of Physics, Long Island University, Brookville, New York 11548, USA}

\date{\today}

\begin{abstract}
  We study the stability of anti--de Sitter (AdS) spacetime to
  spherically symmetric perturbations of a real scalar field in
  general relativity. Further, we work within the context of the ``two
  time framework'' (TTF) approximation, which describes the leading
  nonlinear effects for small amplitude perturbations, and is
  therefore suitable for studying the weakly turbulent instability of
  AdS---including both collapsing and non-collapsing solutions. We
  have previously identified a class of quasi-periodic (QP) solutions
  to the TTF equations, and in this work we analyze their
  stability. We show that there exist several families of QP solutions
  that are stable to linear order, and we argue that these solutions
  represent islands of stability in TTF. We extract the eigenmodes of
  small oscillations about QP solutions, and we use them to predict
  approximate recurrence times for generic non-collapsing initial data
  in the full (non-TTF) system. Alternatively, when sufficient energy
  is driven to high-frequency modes, as occurs for initial data {\em
    far} from a QP solution, the TTF description breaks down as an
  approximation to the full system. Depending on the higher order
  dynamics of the full system, this often signals an imminent collapse
  to a black hole.
\end{abstract}

\maketitle

\section{Introduction}

Of the maximally symmetric solutions of the Einstein equation,
nonlinear stability in general relativity has been proven for both
Minkowski~\cite{christodoulou1993global} and de
Sitter~\cite{friedrich1986existence} spacetimes. In contrast, the
question of stability of anti--de Sitter (AdS) remains formally
open. A key differentiator of AdS as compared to its $\Lambda \ge 0$
counterparts is that with non-dissipating boundary conditions at
infinity, perturbations cannot decay and energy is
conserved~\cite{Friedrich:2014raa}. Based on knowledge of nonlinear
wave propagation in the absence of dissipation, AdS has been
conjectured to be unstable~\cite{dafermos_holzegel,Anderson:2006ax}
(see also~\cite{Holzegel:2013vwa}). This expectation has been
corroborated by numerical simulations---supported by perturbative
arguments---which showed that certain initial configurations evolve to
black holes, no matter how small the initial deviation from AdS was
taken~\cite{Bizon:2011gg}. This work showed, further, that the
eventual gravitational collapse resulted from a turbulent cascade of
energy to high-frequency modes of AdS, mediated by resonant
interactions.

Instability of AdS would have implications for a number of fields,
ranging from potential gravitational instabilities in other
low-dissipation or confining geometries, to thermalization of
conformal field theories (CFTs). In the context of AdS/CFT (within the
regime where general relativity holds in the bulk), the formation and
subsequent evaporation of a black hole in AdS is believed to be dual
to the process of CFT thermalization. The more recent discovery of
initial configurations in the bulk that appear to {\em avoid} black
hole formation~\cite{Buchel:2012uh} was, therefore, somewhat
surprising, as that would indicate non-thermalizing CFT
configurations. This finding led to the identification of several
``islands of stability'' in
AdS~\cite{Dias:2012tq,Maliborski:2013jca,Fodor:2015eia}.

Significant progress towards an analytic understanding of the
dynamics was achieved with the introduction of a powerful perturbative
framework---the two time framework (TTF)---for analyzing small
perturbations of AdS in terms of coupled nonlinear
oscillators~\cite{Balasubramanian:2014cja} (see
also~\cite{Craps:2014vaa}). This framework efficiently captures the
resonant energy-exchange interactions between normal modes, while
effectively ``integrating out'' high-frequency oscillations. TTF led
to the discovery of a pair of quantities---the energy $E$ and particle
number $N$---that are conserved at the leading nonlinear
level~\cite{Craps:2014jwa,Buchel:2014xwa}. These quantities play key
roles in understanding long-term dynamical behaviors, including dual
(direct and inverse) turbulent cascades and non-equipartition of
energy~\cite{Buchel:2014xwa}.

The main purpose of this paper is to establish a large new class of
islands of stability within the TTF approximation. The central stable
equilibria---quasi-periodic~(QP)
solutions~\cite{Balasubramanian:2014cja,Buchel:2014xwa}---form
discrete families, each family itself parametrized by $N$ and $E$. In
this work we (i) construct the families of equilibrium solutions, (ii)
perform a linear stability analysis within TTF showing stability, and
(iii) use the results to understand the long-term behavior of both
collapsing and non-collapsing initial configurations. In particular,
the stability analysis gives rise to a perturbation spectrum that
agrees with and explains ``recurrences''---long-term nearly-periodic
approaches of the configuration to the initial state, first observed
in the Fermi-Pasta-Ulam (FPU) system of coupled
oscillators~\cite{Fermi:1955:SNP}---which were observed numerically in
the full system. Dependence of the families of QP solutions on the
{\em two} continuous parameters $N$ and $E$ extends previously known
one-parameter families (time-periodic
solutions~\cite{Maliborski:2013jca}) and provides a clear connection
between conserved quantities and stable islands.

\subsection{Background}

Following~\cite{Bizon:2011gg}, we restrict analysis to the spherically
symmetric case and four spacetime dimensions. As a proxy for
gravitational degrees of freedom, we take as our model a real massless
scalar field $\phi$ coupled to general relativity. Ignoring gravity, the
scalar field is characterized by normal modes with spatial
wavefunctions
\begin{equation}
  \label{eq:ej}
  e_j(x) = 4\sqrt{\frac{(j+1)(j+2)}{\pi}}\cos^3x\,{}_2F_1\left(-j,3+j;\frac{3}{2};\sin^2x\right),
\end{equation}
and frequencies $\omega_j=2j+3$ ($j=0,1,2,\ldots$).

Since the frequency spectrum is commensurate, nonlinear gravitational
interactions are resonant, and those interactions cause energy to be
readily transferred among the modes\footnote{The frequency spectrum is
  also resonant with a massive scalar, for other spacetime dimensions,
  and in the absence of spherical symmetry.}. Numerical simulations
have shown that for certain initial data, energy is transferred from
low-$j$ to high-$j$ modes---a direct turbulent
cascade~\cite{Bizon:2011gg}. This cascade concentrates the
energy---the high-$j$ modes are more highly peaked in position
space---and eventually leads to black hole formation. The cascade
behavior persists self-similarly as the amplitude $\epsilon$ of the
initial scalar field is decreased, with the time to collapse scaling
as $1/\epsilon^2$ (see, e.g., Fig.~2 of~\cite{Bizon:2011gg}).

In contrast, other initial data seem to avoid
collapse\footnote{Simulations are of finite duration, and the limit
  $\epsilon\to0$ cannot be obtained numerically, so collapse-avoidance
  is a conjecture.} as $\epsilon\to0$. In addition to direct cascades,
these solutions feature {\em inverse} turbulent cascades, which
transfer energy to low-$j$
modes~\cite{Buchel:2013uba,Balasubramanian:2014cja}. Collapse is
avoided if the inverse cascades sufficiently hinder the flow of energy
to high-$j$ modes.

A key observation is that energy cascades and normal mode oscillations
are governed by independent time scales. As $\epsilon\to0$, nonlinear
interactions become weaker---the stress-energy tensor
$T_{ab}^\phi\propto\epsilon^2$, and gravitational self-interactions of
$\phi$ scale as $\epsilon^3$---so the energy transfer time scale is
proportional to $1/\epsilon^2$. Meanwhile, normal mode oscillations
proceed independently of $\epsilon$. This separation of time scales
means we can use multiscale analysis methods to study the slow
mode-mode interactions independently of the fast normal mode
oscillations in the limit $\epsilon\to0$~\cite{Balasubramanian:2014cja}.

We define the ``slow time'' $\tau\equiv t/\epsilon^2$. Over short time
scales the scalar field is well-approximated as a sum over normal
modes. Thus, we take as ansatz $\phi=\epsilon\phi^{(1)}$, with
\begin{equation}\label{eq:phi}
\phi_{(1)}(t,\tau,x) = \sum_{j=0}^{\infty}\left(A_j(\tau)e^{-i \omega_j t} + \bar{A}_j(\tau)e^{i \omega_j t}\right)e_j(x).
\end{equation}
At lowest nonlinear order, we showed that gravitational
self-interactions of $\phi$ are taken into account provided the
coefficients $A_j(\tau)$ satisfy the coupled ordinary differential
equations~\cite{Balasubramanian:2014cja},
\begin{equation}\label{eq:TTF}
-2 i \omega_j \frac{dA_j}{d\tau} = \sum_{klm}\mathcal{S}^{(j)}_{k l m}\bar{A}_k A_l A_m,
\end{equation}
known as the {\em two time framework (TTF) equations}. The TTF
equations were also derived using renormalization group perturbation
methods to re-sum secularly growing terms that arise in ordinary
perturbation theory~\cite{Craps:2014vaa}. Notice that the TTF
equations possess the same scaling symmetry, $A(\tau)\to \epsilon
A(\tau/\epsilon^2)$ seen in the full (non-TTF) system in the limit
$\epsilon\to0$.

The numerical coefficients $\mathcal{S}^{(j)}_{k l m}$ appearing
in~\eqref{eq:TTF} arise from overlap integrals involving the $e_j(x)$,
and they vanish unless $j+k=l+m$.  This fact, together with the
specific form of the equations~\eqref{eq:TTF} (i.e., the lack of terms
such as $\bar{A}_k\bar{A}_lA_m$, etc.), arises because the only
resonances that are present in the system are those such
that~\cite{Craps:2014vaa}
\begin{equation}\label{eq:resonance}
  \omega_j+\omega_k=\omega_l+\omega_m.
\end{equation}
This property is related to a hidden symmetry in
AdS~\cite{Evnin:2015gma}. For further discussion on the absence of
certain resonance channels, see~\cite{Yang:2015jha}.

Within their regime of validity, the TTF equations yield approximate
solutions much more economically than full numerical relativity
simulations~\cite{Balasubramanian:2014cja}. Indeed, a significant
speedup is gained by not modeling the rapid normal-mode
oscillations. Moreover, the TTF approximation improves as
$\epsilon\to0$---a limit that is especially hard to reach in numerical
relativity. Nevertheless, in the same way that finite difference
methods employ a discrete spatial grid, the set of TTF
equations~\eqref{eq:TTF} must in practice be truncated at finite
$j=j_{\text{max}}$ (similar to pseudo-spectral methods). 
Previously, we computed (by performing explicit
integrations on a mode-by-mode basis) the coefficients
$\mathcal{S}^{(j)}_{klm}$ up to
$j_{\text{max}}=47$~\cite{Balasubramanian:2014cja}. We now have closed
form expressions for the coefficients (see
App.~\ref{sec:closedform}) that enable us to work to much larger
$j_{\text{max}}$. We typically set $j_{\text{max}}=200$ in this paper,
which in many cases provides an excellent approximation. In
particular, the recurrence dynamics of non-collapsing solutions are
well-captured.

While useful as a calculational tool, the main power of TTF is
analytic\footnote{See also~\cite{Yang:2015jja} for another recent
  illustration of the power of this approach within general
  relativity.}.  Indeed, in~\cite{Craps:2014jwa,Buchel:2014xwa} it was
uncovered that the TTF equations conserve a total of {\em three}
quantities: The total energy and particle number,
\begin{eqnarray}
  E & \equiv & 4\sum\limits_j{\omega_j^2|A_j|^2},\label{eq:E}\\
  N & \equiv & 4\sum\limits_j{\omega_j|A_j|^2},\label{eq:N}
\end{eqnarray} 
as well as the Hamiltonian\footnote{\label{fn:hamiltonian}As described
  in detail in~\cite{Craps:2014jwa}, the system~\eqref{eq:TTF} in the
  ``origin-time'' spacetime gauge
  of~\cite{Bizon:2011gg,Balasubramanian:2014cja,Craps:2014vaa} is not
  a Hamiltonian system itself. However, in the ``boundary-time'' gauge
  of~\cite{Buchel:2012uh,Buchel:2013uba} the system is Hamiltonian
  with Hamiltonian $H$. Both gauges possess the same conserved
  quantities, so in this paper we shall refer to $H$ as the
  ``Hamiltonian'', despite working in origin-time gauge (for
  comparison with prior numerical simulations). Note also that the
  equivalent expression in~\cite{Buchel:2014xwa} did not include the
  second term in $H$.},
\begin{equation}\label{eq:H}
  H \equiv -\frac{1}{4}\sum_{jklm}\mathcal{S}^{(j)}_{klm}\bar{A}_j\bar{A}_kA_lA_m - \frac{E}{4}\sum_{j}\mathcal{C}_j|A_j|^2,
\end{equation}
where $\mathcal{C}_j$ are additional constants. Conservation laws of
$E$ and $N$ are associated with two $U(1)$ symmetries,
\begin{eqnarray}
  \label{eq:U1-E}A_j(\tau)&\to& A_j(\tau)e^{i\omega_j\theta},\\
  \label{eq:U1-N}A_j(\tau)&\to& A_j(\tau)e^{i\theta},  
\end{eqnarray}
respectively, for $\theta\in\mathbb{R}$ constant; conservation of $H$
is associated with time-translation
symmetry~\cite{Craps:2014jwa}. (The symmetries and associated
conservation laws were first uncovered for the TTF equations that
describe a non-gravitating scalar field in AdS${}_4$, with quartic
self-interaction $V(\phi)=\lambda\phi^4/4!$~\cite{Basu:2014sia}.)
Simultaneous conservation of $E$ and $N$ implies that direct and
inverse turbulent cascades must occur together, and that energy
equipartition is in general not possible~\cite{Buchel:2014xwa}.

Finally, we showed in~\cite{Balasubramanian:2014cja} that the TTF
equations give rise to equilibrium solutions, which are 
QP. That is, each mode amplitude,
\begin{equation}\label{eq:QP-ansatz}
  A_j(\tau) = \alpha_j e^{-i\beta_j\tau},
\end{equation}
with $\beta_j\in\mathbb{R}$. Simulations in TTF and full numerical
relativity both provided evidence for stability of these QP
solutions. The case was then made in~\cite{Buchel:2014xwa} that
general non-collapsing solutions can be treated as perturbations about
associated QP solutions---in other words, QP solutions with the same
$E$ and $N$. As an example application, we studied two-mode initial
data, which exhibits FPU-like~\cite{Fermi:1955:SNP} recurrences over
long time scales. We showed, by interpolating initial data between
two-mode and associated QP, that the recurrence times were only
marginally affected.  We therefore concluded that a proper stability
analysis might predict these times, and QP solutions might provide
anchor points for the ``islands of stability'' in AdS.

\subsection{Summary}

In this paper we present a comprehensive analysis of QP solutions and
their relation to AdS (in)stability. After presenting the algebraic
equations governing QP solutions in
Sec.~\ref{sec:quasiperiodicsolutions}, we show that they extremize $H$
for first order variations holding $E$ and $N$ fixed. We then
numerically map out the space of solutions to the QP equations. This
space can be divided into a number of families of solutions, each one
depending on two continuous parameters, $E$ and $N$.  Because of the
scaling symmetry of the TTF equations, these families are
scale-invariant, so it is often useful to exchange $E$ and $N$ for an
overall scale, and the ratio $T\equiv E/N$---which we identify with
the ``temperature.''

In Sec.~\ref{sec:QPlinear} we perform a linear stability analysis of
QP solutions within TTF. We uncover two 2-dimensional subspaces of
special perturbations: the first corresponds to a pair of generators
of the $U(1)$ symmetries~\eqref{eq:U1-E} and~\eqref{eq:U1-N} of TTF;
the second represents infinitesimal perturbations to nearby QP
solutions with different $E$ and $N$. (We make use of these special
perturbations to generate the continuous families of QP solutions
parametrized by $E$ and $N$ in Sec.~\ref{sec:quasiperiodicsolutions}.)
The remaining perturbations preserve $E$ and $N$, and may be
decomposed into eigenmodes describing small oscillations. The
corresponding eigenvalues determine stability. We present a numerical
method to perform this stability analysis given any particular
background QP solution.

After presenting the framework for analyzing stability, we apply it to
the families of QP solutions identified in
Sec.~\ref{sec:quasiperiodicsolutions}. We argue, by explicitly checking
a large number of QP solutions, that the ``physical'' families---those
that do not depend strongly on the mode cutoff $j_{\text{max}}$ in the
limit $j_{\text{max}}\to\infty$---are all stable. By contrast, QP
solutions that are not members of these families can have unstable
modes.

In Sec~\ref{sec:positionspace} we apply the results of the stability
analysis to understand long-term evolutions. Since $E$ and $N$ are
conserved, motion in phase space is constrained to constant-$(E,N)$
hypersurfaces. Each surface intersects a given stable QP family at
most once, resulting in a discrete collection of QP solutions. If
initial data lies within the $H$-trough around one of these QP
solutions with the same $(E,N)$, then we associate it to that QP
solution. Under evolution the solution is then confined to oscillate
about its associated QP solution. We illustrate, through several
examples, how nonlinear evolutions of initial data within TTF inherit
many of the properties uncovered by the linear stability analysis. In
particular, the linear stability analysis explains nonlinear
recurrences as oscillations about QP solutions, and the eigenmodes
(and combinations thereof) predict the recurrence times. Thus, we
obtain approximate recurrence times without performing any time
integrations. This approach to understanding recurrences is a
generalization (to two conserved quantities) of the $q$-breather
approach to understanding FPU recurrences~\cite{Flach:2005,Flach:2006}.

For evolutions that remain close to stable QP solutions, the energy
spectra remain close to the exponential energy spectra of the QP
solutions. In contrast, solutions that are not close to QP solutions
tend to approach power laws in TTF, consistent with earlier
studies~\cite{Maliborski:2013via}. A power law spectrum contains far
more energy at high-$j$, and when translated to a description
involving spacetime fields at finite $\epsilon$, the energy is far
more concentrated at the origin of AdS; in fact, it fails to even
converge in $j$. When deviations from AdS become large, TTF no longer
applies, and higher-order dynamics take over. It is often the case
that the higher order dynamics rapidly drive collapse once they take
hold~\cite{Bizon:2011gg}; the role of TTF is to indicate whether
this regime is reached.

It is important to keep track of the various levels of levels of
approximation used in this work, so we summarize them here. First, the
TTF equations are taken as an approximation to the full system, valid
in the limit\footnote{For an interesting discussion of when solutions
  of the approximated system might correspond to solutions of the full
  system, see~\cite{Dimitrakopoulos:2015pwa}.}
$\epsilon\to0$. Secondly, we truncate the TTF system at a finite
number $j_{\text{max}}$ of modes. Finally, we perform a linear
stability analysis of QP solutions within the truncated TTF
system. Throughout this work we will address the validity of the
various approximations.

\section{Quasi-periodic solutions}\label{sec:quasiperiodicsolutions}

There is already strong evidence that there are stable equilibrium
solutions---{\em{islands of stability}}---in AdS, namely the
time-periodic solutions~\cite{Bizon:2011gg,Maliborski:2013jca}. These
solutions are nonlinear generalizations of individual normal modes,
with the effect of gravity being to shift the frequency. Such
solutions are moreover realized as solutions to the TTF
system~\eqref{eq:TTF} of the form
\begin{equation}\label{eq:single-mode}
A_j(\tau) = \delta_{jk}A_k(0) e^{\frac{i}{2 \omega_k} \mathcal{S}^{(k)}_{kkk} |A_k(0)|^2 \tau},
\end{equation}
for some fixed mode number $k$. (The analysis
of~\cite{Maliborski:2013jca}, however, is accurate to higher order in
$\epsilon$.) For a given $k$ there exists a 1-parameter family of
solutions, parametrized by $A_k(0)$, or equivalently, the energy $E$.

Inspired by the periodic solutions, we identified
in~\cite{Balasubramanian:2014cja} a {\em much larger class} of {\em
  quasi}-periodic (QP) solutions. Allowing for {\em all} modes to be excited
periodically (but with different periods), we sought solutions of the form
\begin{equation}
  A_j(\tau)=\alpha_je^{-i\beta_j\tau},\nonumber
\end{equation}
with $(\alpha_j,\beta_j) \in \mathbb{C}\times\mathbb{R}$. Such
solutions would have constant energy, $E_j$, in each mode---finely tuned
so that energy flows between modes are perfectly balanced.

Substituting the ansatz above into~\eqref{eq:TTF}, we have
\begin{equation}
  -2\omega_j\beta_j\alpha_je^{-i\beta_j\tau} = \sum_{klm}\mathcal{S}^{(j)}_{klm}\bar{\alpha}_j\alpha_k\alpha_le^{-i(-\beta_k+\beta_l+\beta_m)\tau}.
\end{equation}
We see that the $\tau$-dependence may be canceled from both sides by imposing the condition
\begin{equation}\label{eq:beta}
  \beta_j = \beta_0 + (\beta_1 - \beta_0) j,
\end{equation}
reducing the system to
\begin{equation}\label{eq:QP}
-2 \omega_j \left[\beta_0 + j \left(\beta_1 - \beta_0\right)\right]\alpha_j = \sum_{klm}\mathcal{S}^{(j)}_{k l m}\alpha_k \alpha_l \alpha_m.
\end{equation}
Without loss of generality, henceforth we take $\alpha_j\in\mathbb{R}$ in the
equation above (this represents a choice of initial
time $\tau=0$).  We thus have $j_{\text{max}}+1$ {\em algebraic}
equations for $j_{\text{max}}+3$ unknowns. That is, we have {\em two}
free parameters---one more than the time-periodic solutions---which we
will often take as $E$ and $N$.

\subsection{Extremization of $H$}\label{sec:dH}

Quasi-periodic solutions extremize the Hamiltonian $H$ with respect to
perturbations that preserve $E$ and $N$. To see this, we first
introduce some notation (following~\cite{Craps:2014jwa}). We split the
coefficients
\begin{equation}\label{eq:Sspliting}
  \mathcal{S}^{(j)}_{klm} = \mathcal{S}^{\text{S}}_{jklm} + \mathcal{R}^{\text{A}}_{jk}\left(\delta_{jl}\delta_{km}+\delta_{jm}\delta_{kl}\right),
\end{equation}
where $\mathcal{S}^{\text{S}}_{jklm}$ is symmetric under interchange
of $jk$ with $lm$ (as well as exchange of $j$ with $k$ or $l$ with
$m$). The quantity $\mathcal{R}^{\text{A}}_{jk}$ is antisymmetric and takes the
form
\begin{equation}\label{eq:Rspliting}
  \mathcal{R}^{\text{A}}_{jk}=\mathcal{C}_j\omega^2_k-\mathcal{C}_k\omega^2_j.
\end{equation}
We then define the quantity
\begin{equation}
  V\equiv \sum_{jklm}\mathcal{S}^{(j)}_{klm}\bar{A}_j\bar{A}_kA_lA_m.
\end{equation}
It may be shown that
\begin{equation}
  \frac{\partial V}{\partial\bar{A}_j} = 2\sum_{klm}\mathcal{S}^{\text{S}}_{jklm}\bar{A}_kA_lA_m.
\end{equation}
Thus \eqref{eq:TTF} may be re-written
\begin{equation}\label{eq:TTF-V}
  -2i\omega_j\frac{dA_j}{d\tau} = \frac{1}{2}\frac{\partial V}{\partial \bar{A}_j} + 2\sum_k \mathcal{R}_{jk}^{\text{A}}|A_k|^2A_j,
\end{equation}
or in terms of the Hamiltonian,
\begin{equation}\label{eq:TTF-H}
  i\omega_j\frac{dA_j}{d\tau} = \frac{\partial H}{\partial\bar{A}_j}+2\omega_j^2A_j\sum_k\mathcal{C}_k|A_k|^2.
\end{equation}
Note that the presence of the last term indicates that the system is
not actually Hamiltonian in the ``origin-time'' spacetime gauge in
which we work (see footnote~\ref{fn:hamiltonian}). This term is not
present in the ``boundary-time'' gauge~\cite{Craps:2014jwa}.

Now consider a variation that fixes $E$ and $N$,
\begin{eqnarray}
  \delta H &=& \sum_j\left( \frac{\partial H}{\partial\bar{A}_j}\delta \bar{A}_j + \frac{\partial H}{\partial A_j}\delta A_j  \right)\\
           &=& i\sum_j\omega_j\left( \frac{dA_j}{d\tau} \delta\bar{A}_j - \frac{d\bar{A}_j}{d\tau}\delta A_j\right)\nonumber\\
           && - \frac{1}{2} \sum_{jk}\omega_j^2\left(A_j\delta\bar{A}_j+\bar{A}_j\delta A_j\right)\mathcal{C}_k|A_k|^2\nonumber\\
           &=& i\sum_j\omega_j\left( \frac{dA_j}{d\tau} \delta\bar{A}_j - \frac{d\bar{A}_j}{d\tau}\delta A_j\right) - \frac{\delta E}{2}\sum_k\mathcal{C}_k|A_k|^2.\nonumber
\end{eqnarray}
On the second line we used the TTF equation~\eqref{eq:TTF-H}. On the
last line the final term vanishes for variations that preserve
$E$. For $A_j$ also quasi-periodic, we can now use the
ansatz~\eqref{eq:QP-ansatz} and~\eqref{eq:beta} to simplify the first term,
\begin{eqnarray}\label{eq:HMinimal}
  \delta H &=& \sum_j \omega_j\beta_j\left( A_j\delta \bar{A}_j + \bar{A}_j\delta A_j\right)\nonumber\\
           &=& \sum_j \omega_j \left[\beta_0 + j(\beta_1-\beta_0)\right] \left( A_j\delta \bar{A}_j + \bar{A}_j\delta A_j\right)\nonumber\\
           &=& \frac{1}{8}(\beta_1-\beta_0)\delta E + \frac{1}{4}\left[\beta_0 - \frac{3}{2}(\beta_1-\beta_0)\right] \delta N\nonumber\\
           &=& 0,
\end{eqnarray}
since we fix $E$ and $N$. Thus, QP solutions are critical points of $H$ for
perturbations that fix $E$ and $N$.

\subsection{Families of solutions}\label{sec:qp-solutions-newton}

The QP equations~\eqref{eq:QP} have two free parameters, which must be
fixed prior to solving. But, even after doing so, there remain multiple
solutions because the equations are nonlinear. This gives rise to
multiple {\em families} of QP solutions, each extending over some
range of $E$ and $N$.

We solve the QP equations numerically, following several
approaches described in App.~\ref{sec:QPspace}. As always, the TTF
system is truncated at $j=j_{\text{max}}<\infty$, and the physical
continuum limit corresponds to $j_{\text{max}}\to\infty$. Thus, any QP
solution that depends strongly on $j_{\text{max}}$ must be discarded
as unphysical.

\begin{figure}[tb]
  \centering
    \includegraphics{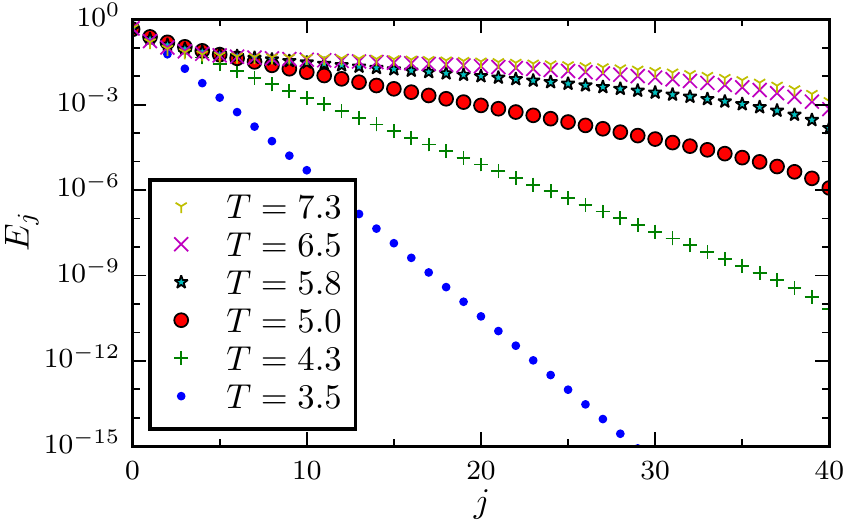}
    \caption{\label{fig:qp-jr0}Energy spectra for several QP solutions
      that were obtained numerically. These solutions are all members
      of the $j_r=0$ family. Here we take $j_{\text{max}} = 40$.}
  
\end{figure}

The simplest way to obtain QP solutions (used
in~\cite{Balasubramanian:2014cja}) is to use a Newton-Raphson method,
which works well if a good initial seed can be chosen. Since we know
that single-mode configurations~\eqref{eq:single-mode} are solutions,
we search for solutions dominated by single modes $j=j_r$, but that
have nonzero energy in the other modes. The energy spectra
$E_j=4\omega_j^2|A_j|^2$ of several such solutions from the $j_r=0$
family are illustrated in Fig.~\ref{fig:qp-jr0}.
Rather than parametrizing the solutions by the continuous parameters
$E$ and $N$ we have labeled the spectra by the temperature
$T=E/N$. The other parameter is simply an overall scale that does not
affect the shape of the curves.

Notice that for small $T$, the energy spectra approach
exponentials. (The minimum temperature for the $j_r=0$ family occurs
in the single-mode limit, with $T_{\text{min}}^{j_r=0} =
E_0/N_0=\omega_0=3$.) For larger $T$ the spectra deform and it becomes
increasingly difficult to obtain solutions using the Newton-Raphson
method. For such cases we can obtain solutions by perturbing known
solutions to different $E$ and $N$ (see
App.~\ref{sec:QPspace}). For the $j_r=0$ family, solutions exist
up to $T=T_{\text{max}}=\omega_{j_{\text{max}}}=2j_{\text{max}}+3$,
which is the maximum possible temperature for the truncated collection
of modes. Such solutions are highly deformed from exponentials---the
maximal solution has all energy in mode $j=j_{\text{max}}$---and are
not physical because of the dependence on mode truncation.  Requiring
$T\ll T_{\text{max}}$ will select for physical configurations, and,
with this restriction, the physically relevant spectra are all nearly
exponential. Extrapolating to the continuum limit
$j_{\text{max}}\to\infty$---where by definition there are no
unphysical solutions---we expect all $j_r=0$ solutions to have
nearly-exponential spectra (for any $T$).

\begin{figure}[tb]
  \centering
    \includegraphics{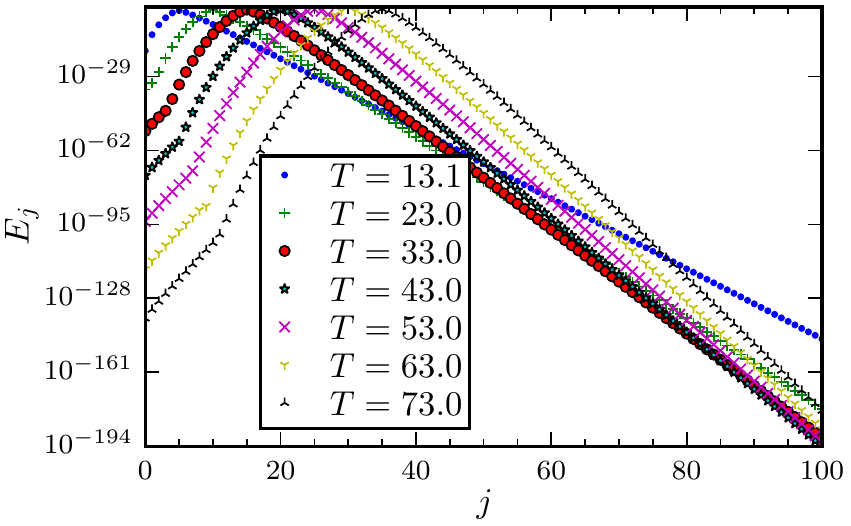}
    \caption{\label{fig:qp-jrMulti}Energy spectra of QP solutions from
      several discrete families (with different $j_r$). The
      temperature in each case is very close to $\omega_{j_r}$, as the
      QP solutions shown here are very close to single-mode
      solutions. ($j_{\text{max}} = 100$)}
  
\end{figure}

\begin{figure}[tb]
  \centering
  \includegraphics{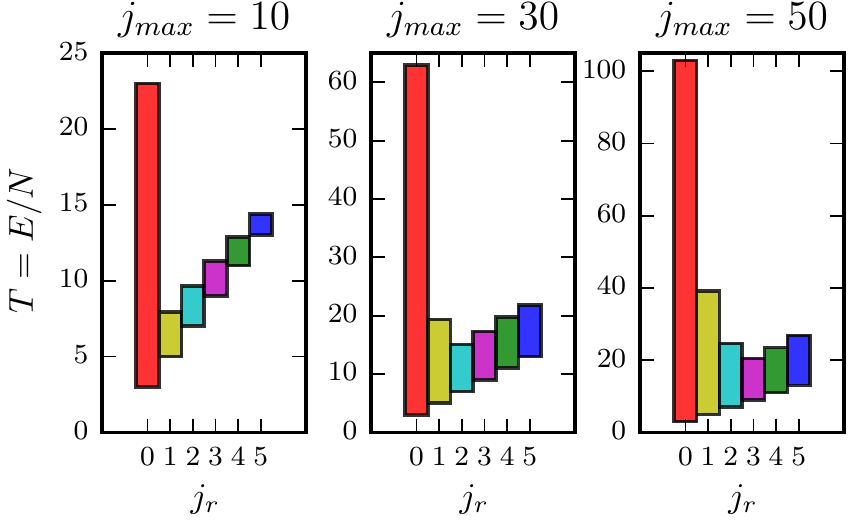}
  \caption{\label{fig:SizeFamilies}The domain of existence of QP
    families for $j_r \in \{0,1,2,3,4,5\}$ and $j_\text{max} \in
    \{10,30,50\}$. For $j_r = 0$, the family is defined in the full
    domain $\left[3,2 j_\text{max} +3 \right]$. Note that the bounds
    of the vertical axis increases with $j_\text{max}$.}
\end{figure}

In Fig.~\ref{fig:qp-jrMulti} we plot the spectra of QP solutions from
families with various $j_r>0$. Each solution is peaked at $j=j_r$,
and decays exponentially to both sides (with slight deformation for
$j<j_r$).
As $j_r$ increases, so does the minimum temperature
$T^{j_r}_{\text{min}}=\omega_{j_r}$ of the respective QP family. We
find that the $j_r>0$ families do not extend in temperature all the
way to $T_{\text{max}}$ (in contrast to the $j_r=0$ case), but that
the range of temperatures increases with $j_{\text{max}}$ (see
Fig.~\ref{fig:SizeFamilies}). In the $j_{\text{max}}\to\infty$ limit
it is not clear whether the families have a finite or infinite extent.
  
\begin{figure}[tb]
  \centering
    \includegraphics{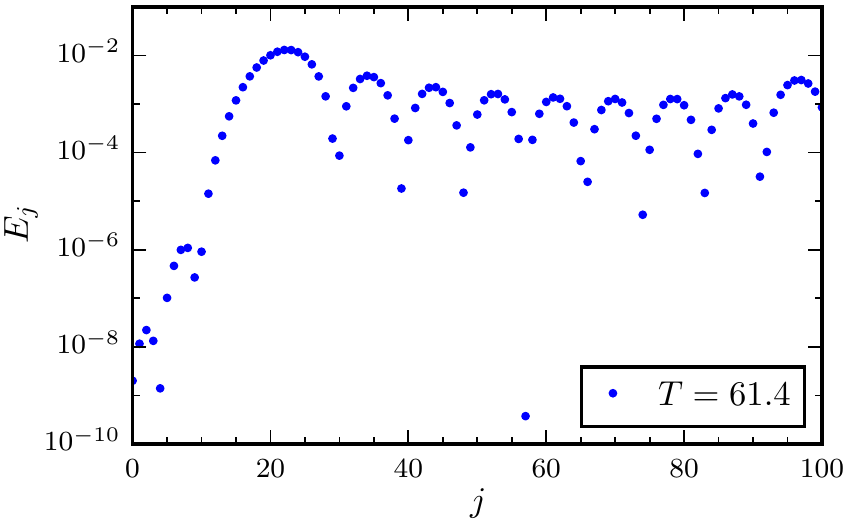}
    \caption{\label{fig:qp-unstable}QP solution not smoothly connected
      to single-mode solution. ($j_{\text{max}} = 100$)}
  
\end{figure}

It is possible to construct additional QP families. For example, the
resonance condition~\eqref{eq:resonance} implies that if only
even-numbered modes are excited initially, they will never excite
odd-numbered modes. In this case QP solutions can be found that are
similar to those of Fig.~\ref{fig:qp-jr0}, but skipping every other
mode. Finally, there are solutions that have considerable energy in
high-$j$ modes that do not appear to connect to the families above
(see Fig.~\ref{fig:qp-unstable}). These latter solutions are clearly
dependent on mode-truncation, so we discard them as unphysical.

\section{Stability of quasi-periodic solutions}\label{sec:QPlinear}

In~\cite{Balasubramanian:2014cja}, we numerically tested the stability
of several QP solutions in the $j_r=0$ family within the full
(non-TTF) theory. For the duration of the simulations, perturbations
oscillated about the QP solutions over time scales long compared to
the AdS crossing time.

To address stability more systematically, in this section we undertake
a linear stability analysis of QP solutions within TTF. In
Sec.~\ref{sec:linpert} we linearize the equations~\eqref{eq:TTF} about
an arbitrary background QP solution. We show that through an
appropriate change of variables, the time-dependence (resulting from a
time-dependent background solution) can be eliminated, leaving an
autonomous system of the form,
\begin{equation}
  \frac{d\boldsymbol{x}}{d\tau} = \boldsymbol{Ax}.\nonumber
\end{equation}
The matrix $\boldsymbol{A}$ depends on the background QP solution, and
is {\em independent of time $\tau$}. The problem of solving in time
for the perturbation vector $\boldsymbol{x}$ is, therefore, equivalent
to that of diagonalizing $\boldsymbol{A}$.

In Sec.~\ref{sec:specialsolutions} we identify special solutions
(infinitesimal $U(1)$ symmetry transformations and perturbations to
other QP solutions) unrelated to stability, and in
Sec.~\ref{sec:generalsolution} we outline the numerical procedure for
finding the remaining eigenvalues of $\boldsymbol{A}$. Finally, in
Sec.~\ref{sec:stabresults} we apply this approach to study the
stability of the families of QP solutions identified in
Sec.~\ref{sec:qp-solutions-newton}. We sample a large number of
solutions within the ``physical'' families, and find that they are all
Lyapunov stable---initially small perturbations remain small, but they
do not decay to zero (see Sec.~23 of~\cite{Arnold:ODEs}). In
Sec.~\ref{sec:stabham} we comment on nonlinear stability for
finite-sized perturbations.

Throughout this paper, we work in the origin-time spacetime gauge (see
footnote~\ref{fn:hamiltonian}). We note that the stability analysis
would go through nearly identically in the boundary-time gauge, where
the system is truly Hamiltonian. The change of gauge only contributes
a time-dependent phase shift to the TTF coefficients $A_j(\tau)$, and
our stability results hold in both gauges.

\subsection{Linearized equations}\label{sec:linpert}

Consider a perturbed QP solution,
\begin{equation}
\label{eq:perturbation}
  A_j(\tau) = A_j^{\text{QP}}(\tau) + \xi_j(\tau),
\end{equation}
where $A_j^{\text{QP}}(\tau) = \alpha_je^{-i\beta_j\tau}$, with
$\{ \alpha_j, \beta_j \} \in \mathbb{R}$. Substituting
into~\eqref{eq:TTF} and keeping terms to first order in $\xi_j$, we have
\begin{eqnarray}\label{eq:xiode}
\frac{d \xi_j}{d\tau} & = &\frac{i}{2\omega_j}\sum_{klm}\mathcal{S}^{(j)}_{k l m}\left[\bar{\xi}_k \alpha_l \alpha_m e^{-i \tau (\beta_l + \beta_m)}\right. \\
                    &&+ \left. \bar{\alpha}_k \xi_l \alpha_m e^{-i \tau (\beta_m - \beta_k)}+ \bar{\alpha}_k \alpha_l\xi_m e^{-i \tau (\beta_l - \beta_k)}\right].\nonumber
\end{eqnarray}
Since the background QP solution has quasi-periodic time-dependence,
so do the coefficients of this equation. If the coefficients were in
fact {\em periodic} one could have applied the Floquet theory to
obtain the general solution to~\eqref{eq:xiode} in terms of
eigenmodes, and thereby determine stability (see Sec.~28
of~\cite{Arnold:ODEs}). (In fact, by tweaking the values of $\beta_0$
and $\beta_1$ so that they are rational multiples of each other,
periodicity can be achieved, although the period might be quite long.)
The Floquet approach requires numerical integrations over one period
to identify the eigenmodes, which is somewhat tedious, but works
generically for periodic systems.

For our TTF system, however, the analysis simplifies due to the
resonance condition. First, factor out the background time-dependence
in each perturbative mode to define new variables,
\begin{equation}
\label{eq:nobeta}
  \xi_j(\tau) = \chi_j(\tau) e^{-i\beta_j\tau}.
\end{equation}
This gives rise to the {\em autonomous} equations
\begin{eqnarray}\label{eq:LinearStabEquationsWithConjugate}
  \frac{d\chi_j}{d\tau} &=& i\beta_j\chi_j \\
  &&+\frac{i}{2\omega_j}\sum_{klm}\mathcal{S}^{(j)}_{klm}\left(\bar{\chi}_k\alpha_l\alpha_m+\bar{\alpha}_k\chi_l\alpha_m + \bar{\alpha}_k\alpha_l\chi_m\right).\nonumber
\end{eqnarray}
These equations contain complex conjugations of $\chi_j$ and are
therefore not linear over $\mathbb{C}$. To obtain a linear system,
split $\chi_j$ into its real and imaginary parts,
\begin{equation}
\chi_j(\tau) =  u_j(\tau) + i  v_j(\tau).
\end{equation}
The system is now reduced to
\begin{eqnarray}
  \label{eq:gamma}\frac{d  u_j}{d\tau} &=& -\beta_j  v_j \\
                         &&- \frac{1}{2 \omega_j}\sum_{klm}\mathcal{S}^{(j)}_{k l m}\left(-\alpha_l \alpha_m  v_k + \alpha_m \alpha_k  v_l +\alpha_l \alpha_k  v_m\right),\nonumber\\
  \label{eq:delta}\frac{d  v_j}{d\tau} &=& \beta_j  u_j \\
                         &&+ \frac{1}{2 \omega_j}\sum_{klm}\mathcal{S}^{(j)}_{k l m}\left(\alpha_l \alpha_m  u_k + \alpha_m \alpha_k  u_l  +\alpha_l \alpha_k  u_m\right).\nonumber
\end{eqnarray}

It can be shown that the equations \eqref{eq:gamma}--\eqref{eq:delta}
conserve the {\em linearized} energy, particle number, and
Hamiltonian,
\begin{eqnarray}
  \delta E &=& 8\sum_j\omega_j^2\alpha_j u_j,\\
  \delta N &=& 8\sum_j\omega_j\alpha_j u_j,\\
  \delta H &=& \frac{1}{8}\left(\beta_1-\beta_0-4\sum_j\mathcal{C}_j\alpha_j^2\right)\delta E \nonumber\\
           &&+ \frac{1}{8}\left(5\beta_0-3\beta_1\right)\delta N.
\end{eqnarray}

\subsection{Special solutions}\label{sec:specialsolutions}

\subsubsection{$U(1)$ symmetry transformations}

Recall that the TTF equations are invariant under two $U(1)$
symmetries~\eqref{eq:U1-E}--\eqref{eq:U1-N},
\begin{eqnarray*}
  A_j(\tau)&\to& A_j(\tau)e^{i\omega_j\theta},\\
  A_j(\tau)&\to& A_j(\tau)e^{i\theta},
\end{eqnarray*}
for $\theta\in\mathbb{R}$ constant. Off of QP solutions, infinitesimal
$U(1)$ transformations take the form
\begin{eqnarray}
  \label{eq:U(1)-E-inf}\left(
  \begin{array}{c}
     u_j \\  v_j
  \end{array}
  \right) &\to& \left(
                \begin{array}{c}
                  0 \\ \omega_j\alpha_j\theta
                \end{array}
  \right), \\
  \label{eq:U(1)-N-inf}\left(
  \begin{array}{c}
     u_j \\  v_j
  \end{array}
  \right) &\to& \left(
                \begin{array}{c}
                  0 \\ \alpha_j\theta
                \end{array}
  \right),
\end{eqnarray}
respectively.

It is straightforward to check that these perturbations
satisfy~\eqref{eq:gamma}--\eqref{eq:delta}. Indeed,~\eqref{eq:delta}
holds trivially, while~\eqref{eq:gamma} holds because of the resonance
condition~\eqref{eq:resonance} [in the case of~\eqref{eq:U(1)-E-inf}]
and the QP equation~\eqref{eq:QP}.

\subsubsection{Perturbations to nearby QP solutions}\label{sec:nearbyQP}

Consider now a perturbation from a QP solution to another QP solution,
\begin{equation}\label{eq:nearby-QP-ansatz}
  \alpha_je^{-i\beta_j\tau}\to (\alpha_j+\delta\alpha_j)e^{-i(\beta_j+\delta\beta_j)\tau}.
\end{equation}
The new QP solution is required to satisfy the QP
equation~\eqref{eq:QP} as well. To first order in the perturbation,
this requirement takes the form
\begin{eqnarray}\label{eq:nearby-QP}
  &&-2\omega_j\left(\alpha_j\delta\beta_j+\beta_j\delta\alpha_j\right)\\
  &=&\sum_{klm}\mathcal{S}^{(j)}_{klm}\left(\alpha_l\alpha_m\delta\alpha_k+\alpha_k\alpha_m\delta\alpha_l+\alpha_k\alpha_l\delta\alpha_m\right),\nonumber
\end{eqnarray}
and the condition~\eqref{eq:beta} implies either of
\begin{eqnarray}
  \label{eq:nearby-QP-case1}\delta\beta_j&\to&\omega_j\theta,\\
  \label{eq:nearby-QP-case2}\delta\beta_j&\to&\theta,
\end{eqnarray}
for $\theta\in\mathbb{R}$.

The infinitesimal version of the
perturbation~\eqref{eq:nearby-QP-ansatz} is
\begin{equation}\label{eq:inf-QP-case2}
  \left(
  \begin{array}{c}
     u_j \\  v_j
  \end{array}
  \right) \to \left(
                \begin{array}{c}
                  \delta\alpha_j \\ -\alpha_j\tau\delta\beta_j
                \end{array}
  \right).
\end{equation}
Using this mapping it is easily checked that~\eqref{eq:nearby-QP}
is identical to~\eqref{eq:delta}, and that~\eqref{eq:gamma} holds for
both cases~\eqref{eq:nearby-QP-case1} and~\eqref{eq:nearby-QP-case2}.

Perturbations~\eqref{eq:nearby-QP-case1}
and~\eqref{eq:nearby-QP-case2} represent a 2-parameter family of
solutions to the linearized equations. This family can be
re-parametrized in terms of $\delta E$ and $\delta N$, allowing for
the families of QP solutions in Sec.~\ref{sec:qp-solutions-newton} to
be fully obtained as orbits of these perturbations (see
App.~\ref{sec:QPspace}).

\medskip

Together, infinitesimal $U(1)$ transformations and infinitesimal
perturbations to nearby QP solutions form two 2-dimensional
generalized eigenspaces (with eigenvalue 0) of the matrix
$\boldsymbol{A}$ representing the linear system (see below). Indeed,
the action of $\boldsymbol{A}$ on a perturbation of the
form~\eqref{eq:nearby-QP-case1} gives a $U(1)$
transformation~\eqref{eq:U(1)-E-inf}, and a subsequent action of
$\boldsymbol{A}$ gives 0. [Similarly,
\eqref{eq:nearby-QP-case2}$\xrightarrow{\boldsymbol{A}}$\eqref{eq:U(1)-N-inf}$\xrightarrow{\boldsymbol{A}}0$.]
A 2-dimensional generalized eigenspace does give rise to linear growth
in the solution [see~\eqref{eq:inf-QP-case2}], but this growth is not
relevant to the question of stability since it is simply an
infinitesimal perturbation to another equilibrium
solution~\eqref{eq:nearby-QP-ansatz}.

\subsection{General solution technique}\label{sec:generalsolution}

It is convenient to express \eqref{eq:gamma}--\eqref{eq:delta} in
matrix form. Defining
\begin{equation}
  \boldsymbol{x} = \left( 
    \begin{array}{c}
      \left( u_j\right)\\
      \left( v_j\right)
    \end{array}
  \right),
\end{equation}
the perturbative equations take the form
\begin{equation}\label{eq:matrixform}
  \frac{d\boldsymbol{x}}{d\tau} = \boldsymbol{Ax},
\end{equation}
where $\boldsymbol{A}$ is a
$(2j_{\text{max}}+2)\times(2j_{\text{max}}+2)$ constant real
matrix. We now complexify the equation and put $\boldsymbol{A}$
in Jordan form, taking real solutions in the end.

In general, the background QP solution, and hence the matrix
$\boldsymbol{A}$, are known only numerically. This is problematic since
the Jordan decomposition is numerically ill-conditioned---if
$\boldsymbol{A}$ has multiple eigenvalues, small errors in
$\boldsymbol{A}$ can lead to large errors in its Jordan
form. In particular, we know from the previous subsection that
$\boldsymbol{A}$ has two generalized eigenspaces of dimension 2, which
can be misidentified as distinct 1-dimensional eigenspaces.

In contrast to the Jordan decomposition, the Schur decomposition is
well-conditioned numerically and continuous in the matrix elements.
We therefore perform a Schur decomposition of $\boldsymbol{A}$,
\begin{equation}
  \boldsymbol{S} = \boldsymbol{U}^{-1}\boldsymbol{A}\boldsymbol{U}.
\end{equation}
Here $\boldsymbol{U}$ is unitary, and the matrix $\boldsymbol{S}$ is
upper triangular with eigenvalues along its diagonal. The known
generalized eigenspaces of $\boldsymbol{A}$ have eigenvalue
0. Numerically, however, these may deviate slightly from zero. We also
find that, generically, all other eigenvalues are well-separated from
0. So, to correct the errors in the generalized eigenspaces, we round
off all infinitesimal diagonal components of $\boldsymbol{S}$ to 0,
and denote this new matrix $\tilde{\boldsymbol{S}}$.

Finally, we take the Jordan decomposition of $\tilde{\boldsymbol{S}}$,
\begin{equation}
  \boldsymbol{J} = \boldsymbol{P}^{-1}\tilde{\boldsymbol{S}}\boldsymbol{P}.
\end{equation}
The matrix $\boldsymbol{J}$ always contains the expected pair of
$2\times2$ Jordan blocks. Aside from these, we found that the Jordan
form $\boldsymbol{J}$ was always diagonal. We denote the additional
$\left(2j_{\text{max}}-2\right)$ eigenvalues by $\lambda_n$, and the
associated eigenvectors of $\boldsymbol{A}$ (the column vectors of
$\boldsymbol{UP}$) by $\hat{\boldsymbol{e}}_n$.

To obtain the time-evolution of a linearized perturbation of a QP
background (of the same $E$ and $N$) one must project initial data
onto the eigenvectors $\{\hat{\boldsymbol{e}}_n\}$. Each of these
eigenvectors then evolves independently as
$e^{\lambda_n\tau}\hat{e}_n$. If the initial data is real then a real
solution is guaranteed.

There are relationships between the eigenvalues of
$\boldsymbol{A}$. Since $\boldsymbol{A}$ is real, if $\lambda$ is an
eigenvalue, then so must be $\bar{\lambda}$. Also, since
$\boldsymbol{A}$ is of the form
$\begin{pmatrix*}[c] 0 & -\boldsymbol{C
  } \\
  \boldsymbol{D} & 0
\end{pmatrix*}$
[see ~\eqref{eq:gamma}--\eqref{eq:delta}],
$\boldsymbol{A}^2 = \begin{pmatrix*}[c]
  -\boldsymbol{C D} & 0 \\
  0 & -\boldsymbol{C D}
\end{pmatrix*}$,
and each eigenvalue of $\boldsymbol{A}^2$ occurs twice. Since these
eigenvalues are the squares of eigenvalues of $\boldsymbol{A}$, and
the eigenvalues of $\boldsymbol{A}$ (excepting 0) are generically
non-degenerate, if $\lambda$ is an eigenvalue of $\boldsymbol{A}$,
then so must be $-\lambda$. In sum,
$(\lambda,-\lambda,\bar{\lambda},-\bar{\lambda})$ must all be
eigenvalues.

These properties of the eigenvalues are also characteristic of
symplectic flows and a Hamiltonian structure. While our system is not
Hamiltonian (in the ``origin-time'' gauge used
here~\cite{Craps:2014jwa}), the pattern of eigenvalues is nevertheless
preserved. In particular, for any decaying mode there exists a
corresponding growing mode, so the best one can hope to achieve in
terms of stability is Lyapunov stability. In this case, all modes have
harmonic time-dependence with no growth or decay---i.e., purely
imaginary $\lambda$.

\subsection{Results}\label{sec:stabresults}

We applied the above analysis to a sampling of the QP solutions
described in Sec.~\ref{sec:qp-solutions-newton}. In almost all cases
we found that all of the eigenvalues $\lambda_n$ were purely
imaginary, implying stability. The only unstable QP solutions were
those previously deemed ``unphysical'', as in
Fig.~\ref{fig:qp-unstable}, and in these cases only a small number of
eigenvalues had nonzero real part. Therefore, we expect that {\em all} of
the ``physical'' QP solutions are stable. For these stable solutions,
we denote the conjugate eigenvalues by using negative indices,
$\lambda_{-n}=-\lambda_n$.

\begin{figure}[tb]
  \centering
  \includegraphics{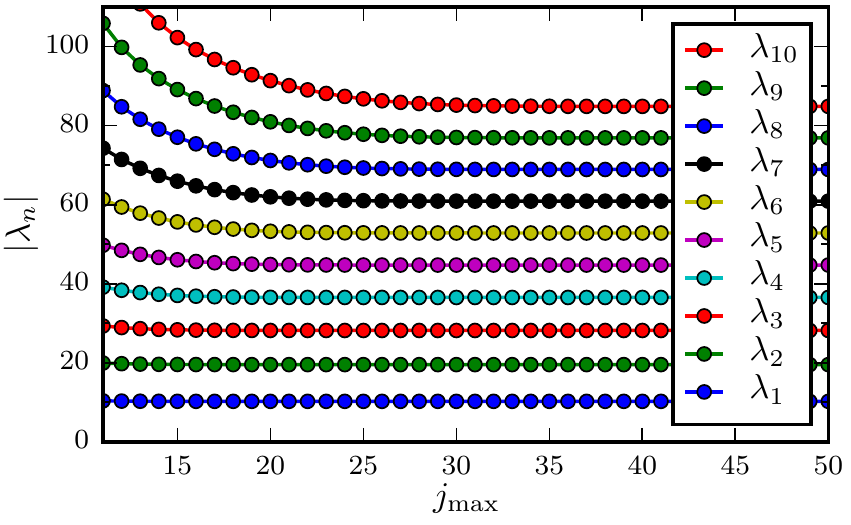}
  \caption{\label{fig:eigenfrequenciesVSjmax}Dependence of
    eigenfrequencies on truncation $j_{\text{max}}$. We plot the ten lowest
    eigenfrequencies for the QP solution with $T = 3.75$ and
    $E = 8$.  After decreasing noticeably up to $j_\text{max}\approx25$,
    eigenfrequencies approach asymptotic values.}
\end{figure}

We studied the dependence of the eigenvalues on $j_{\text{max}}$. As
$j_{\text{max}}$ is increased by 1, a pair of higher frequency
(conjugate) eigenmodes is introduced, while (the norms of the)
existing eigenvalues are shifted slightly lower. In the continuum
limit $j_{\text{max}}\to\infty$, the eigenvalues appear to approach
asymptotic values (see Fig.~\ref{fig:eigenfrequenciesVSjmax}). In that
sense, the behavior of the low-frequency modes is robust to
mode-truncation.

Of particular interest is whether the frequency spectrum is itself
resonant, as this may imply chaotic dynamics at the nonlinear
level. In fact, at high frequencies the separation between subsequent
eigenmodes $\lambda_n$ approaches a constant value as
\begin{equation}\label{eq:asympspectrum}
i\lambda_n = C_1 + C_2n + O\left(\frac{1}{n}\right),
\end{equation}
where $C_1$ and $C_2$ are constants depending on the particular QP
solution\footnote{For single-mode solutions~\eqref{eq:single-mode}
  with $j_r=k$, the eigenvalues may be computed analytically,
  \begin{equation}
    i\lambda_n = \left(\frac{2}{\omega_k}\mathcal{S}^{(k)}_{kkk} -\frac{1}{\omega_n}\mathcal{S}^{(n)}_{knk}\right)[A_k(0)]^2,
  \end{equation}
  provided $A_k(0)\in\mathbb{R}$ (consistent with previous results
  showing stability~\cite{Maliborski:2013jca}). From this spectrum,
  the expansion~\eqref{eq:asympspectrum} may be checked explicitly,
  and the constants $C_1$ and $C_2$ computed.}.  Thus, the
high-frequency part of the spectrum approaches a commensurate spectrum
{\em only asymptotically} in $n$. (We will see later that $C_2$ is
closely related to the recurrence time for non-collapsing solutions.)

\begin{figure}[tb]
  \centering
  \includegraphics{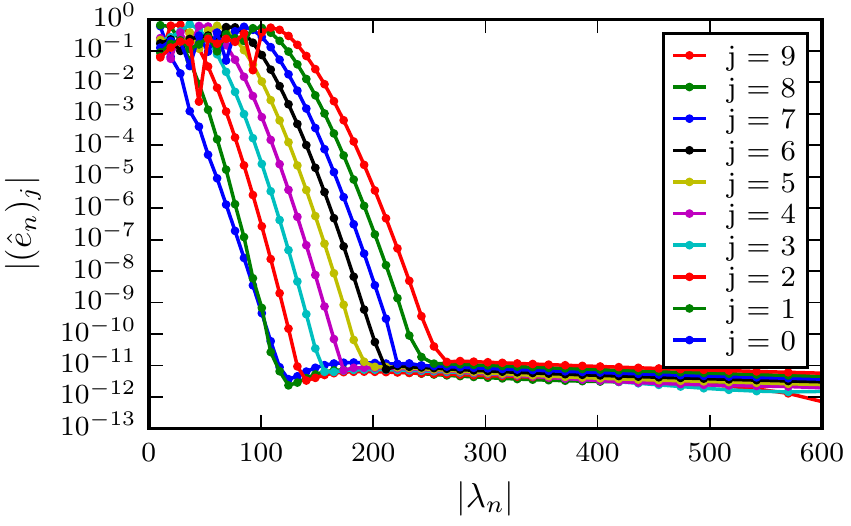}
  \caption{\label{fig:eigenmodesComponents}For the QP solution with
    $T = 3.75$ and $E = 8$, we plot the magnitude of the components
    $(\hat{\boldsymbol{e}}_n)_j$ of the linearized eigenvectors, as a
    function of eigenfrequency $i\lambda_n$. We see that low-frequency
    eigenvectors trigger low-$j$ modes. ($j_{\text{max}}=50$)}
\end{figure}

For perturbations of QP solutions, it is also instructive to examine
the overlap between the original normal modes of AdS ($j$-modes) and
the QP eigenmodes ($n$-modes). The generic solution
to~\eqref{eq:matrixform} is
\begin{equation}
\boldsymbol{x}(\tau) = \sum_n c_n e^{\lambda_n\tau}\hat{\boldsymbol{e}}_n,
\end{equation}
where the $c_n$ are constants. For each $j$,
Fig.~\ref{fig:eigenmodesComponents} plots the components
$(\hat{\boldsymbol{e}}_n)_j$ as a function of eigenfrequency
$i\lambda_n$. This shows that for initial perturbations consisting of
low-$j$ modes, low frequency $n$-modes are excited. Conversely,
low-$n$ eigenmodes excite low-$j$ normal modes most strongly. This
observation explains why low-$j$ modes are typically seen to oscillate
with the lowest frequencies (see, e.g.,
Figs.~\ref{fig:FrequenciesTable} and~\ref{fig:2modeEE-evolution}).

\subsection{Nonlinear stability}\label{sec:stabham}

The linearized analysis above provides useful information and
intuition for finite-sized deviations from QP solutions as well. Since
$E$ and $N$ are conserved quantities, motion in phase space is
constrained to constant--$(E,N)$ hypersurfaces. Each of these
surfaces, in turn, intersects the families of stable QP solutions at
most once each. Thus, given the temperature $T$ of the initial data,
there is a finite number of potentially-relevant QP solutions, and
these can be determined from Fig.~\ref{fig:SizeFamilies}.

Within a given $(E,N)$-surface, we know that the QP solutions
extremize the Hamiltonian $H$, which is also conserved in time. In
App.~\ref{sec:ddH} we show that, in fact, the {\em stable} QP
solutions {\em minimize} $H$. Since $H$ is conserved in time, the size
of the surrounding valley in $H$ determines the size of the island of
stability of the QP solution. One could in principle check to see
whether given initial data lie within one of the valleys, in which
case they would remain near the QP solution indefinitely (within
TTF). As we will see in the following section, nonlinear solutions
often depend closely on the properties (such as the spectrum
$\{\lambda_n\}$) of linearized perturbations about QP solutions.

\begin{figure}[tb]
  \centering
  \includegraphics{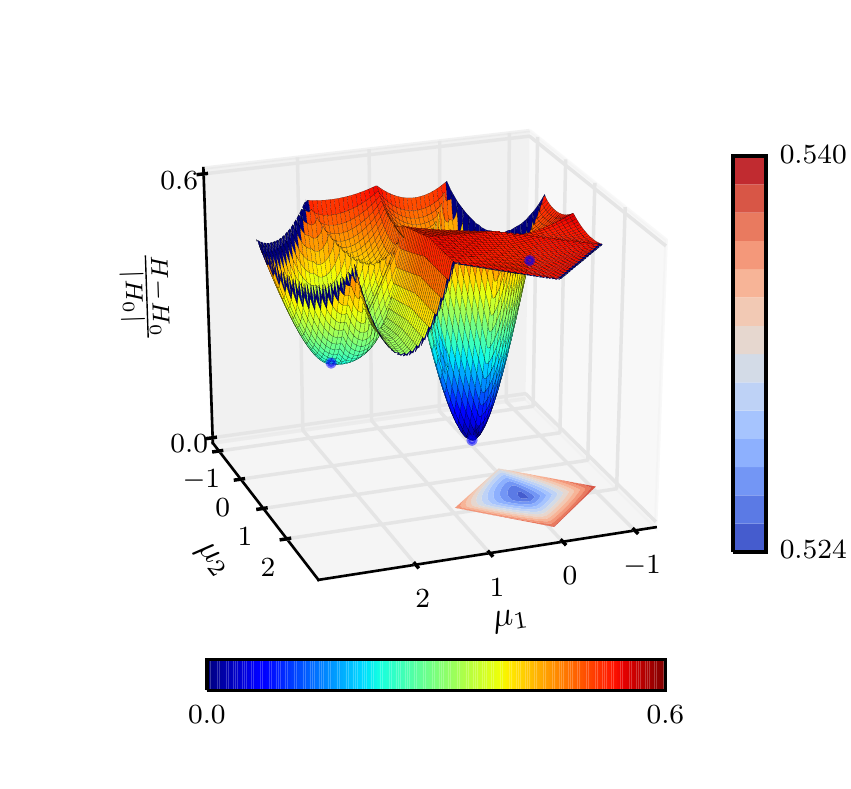}
  \caption{\label{fig:Hplot}Hamiltonian plotted as a function of two
    parameters, $\mu_1$ and $\mu_2$, that interpolate [within
    a constant--$(E,N)$ surface] between 3 QP solutions with
    $T=7.1$. The three QP solutions (blue dots) are members of the
    $j_r=0,1,2$ families. The normalization $H_0$ is the value of $H$
    for the $j_r=0$ QP solution. Note that the ridges result from
    choosing a non-smooth interpolation.}
\end{figure}
It is instructive to visualize the minima of $H$. In
Fig.~\ref{fig:Hplot} we plot the value of $H$ over a two-dimensional
slice of a constant--$(E,N)$ surface. The slice was chosen to pass
through three minima, corresponding to stable QP solutions. Note,
however, that the full problem has a large number of dimensions, with
$(2j_{\text{max}})$-real-dimensional constant--$(E,N)$
hypersurfaces. Moreover, the continuum limit takes
$j_{\text{max}}\to\infty$. While a valley within a finite-dimensional
space must have a finite size, it is possible that this size
asymptotes to zero as $j_{\text{max}}\to\infty$.

\section{Application to AdS (in)stability}\label{sec:positionspace}

We now return to our main questions: for a self-gravitating scalar
field in AdS, how can we predict which initial data will collapse in
the limit of $\epsilon\to0$? How do recurrences arise? How does
collapse in the full Einstein-scalar system connect to behavior in
TTF?

The TTF equations provide a good approximation if the amplitude of the
AdS perturbation is small so that normal-mode oscillation time scales
and mode-mode energy transfer time scales decouple. The approximation,
therefore, always breaks down prior to black hole formation. Knowledge
of this fact alone, however, indicates that a great deal of energy has
transferred to high-$j$ modes, and in many cases, subsequent evolution
will lead to collapse\footnote{In general relativity, collapse usually
  occurs once TTF breaks down~\cite{Bizon:2011gg}, while in
  Gauss-Bonnet gravity it can be averted~\cite{Deppe:2014oua} because
  of a radius gap for black hole formation. This holds despite both
  theories having identical TTF equations~\cite{Buchel:2014dba}.}.

For small, but finite, perturbations, a simple criterion for checking
whether TTF has broken down is to evaluate spacetime quantities
$(\phi,g_{ab})$ from $\{A_j(\tau)\}$ and check for black holes (i.e.,
check whether the metric quantity $A$ of~\cite{Bizon:2011gg} vanishes
at any point, or whether the energy in the scalar field satisfies the
``hoop-conjecture''~\cite{MTW}). Likewise, one could check whether
$(\partial_t\phi)^2$ becomes large. We will refer to the blow-up of
spacetime fields as ``collapse'' in the following, recognizing also
that higher-order dynamics will play a role.

To study stability, one is interested in the $\epsilon\to0$ limit. In
this case, for collapse to occur, spacetime quantities must continue
to be large in this limit. Recalling that spacetime quantities are
generally given as mode sums multiplied by powers of $\epsilon$, it
would be necessary for these mode sums to diverge to see an indication
of collapse. (We are supposing that $j_{\text{max}}\to\infty$ for this
discussion.) For example, $\phi=\epsilon\phi^{(1)}$, with $\phi^{(1)}$
given by~\eqref{eq:phi}, so the only way for $\phi$ to become large in
the $\epsilon\to0$ limit is for the sum~\eqref{eq:phi} to diverge. In
this scenario it is possible to have perfectly well-defined TTF
evolution, but with spacetime quantities ill-defined for any value of
$\epsilon$.

For exponential spectra, $A_j\sim e^{-\mu j}$, sums such
as~\eqref{eq:phi} always converge. But for power laws,
$A_j\sim (1+j)^{-\alpha}$, this is not the case. Indeed, at the
origin, where the mode functions peak,
\begin{equation}
  e_j(0) = \frac{4 \sqrt{j^2+3 j+2}}{\sqrt{\pi }} = O(j).
\end{equation}
So, for example,
\begin{eqnarray}
  \phi(t,0) &=& \epsilon \sum_{j=0}^\infty \left(A_j(\tau)e^{-i\omega_jt}+\bar{A}_j(\tau)e^{i\omega_jt}\right)e_j(0) \nonumber\\
            &\sim& \epsilon \sum_j (1+j)^{-\alpha}\times O(j),
\end{eqnarray}
thus for $\alpha < 1$, $\phi$ is UV-divergent. (We have here assumed
that the phases do not cause a cancellation.)  Other quantities, such
as the metric variable $A$, are even more divergent. We therefore
propose that the large-$j$ asymptotic behavior determines whether
black hole collapse can occur in the $\epsilon\to 0$ limit. (See
also~\cite{Basu:2015efa} for further discussion on this point.)

Connecting to our study of QP solutions, the picture that emerges with
regard to collapse is as follows. QP solutions that have
asymptotically exponential tails will not collapse because they are
equilibria and have well-behaved associated spacetime
quantities. Initial data sufficiently {\em close} to a stable QP
solution (with the same $E$ and $N$) will also not collapse because
the solution will simply oscillate around that QP solution, and its
high-$j$ tail will be close to the QP tail. Initial data that
oscillate about a stable QP solution, but whose oscillations are quite
large {\em can} collapse if the oscillation passes through a power law
that causes the TTF description to break down. Finally, initial data
that do not oscillate about QP solutions can attain a wider range of
configurations, and, as we will confirm, tend to approach power laws
and collapse (in AdS${}_4$).

In the following, we will examine several example solutions within
TTF, both non-collapsing and collapsing. For the non-collapsing
examples, our approach is to identify the closest stable QP solution,
and relate the observed dynamics to the linearized analysis. We find
that the linearized eigenfrequencies $\{\lambda_n\}$ (and combinations
thereof) do a remarkable job of approximating the recurrence times,
even for large perturbations. It should be kept in mind that, while
the physically relevant limit takes $j_{\text{max}}\to\infty$, all
simulations are by necessity performed at finite
$j_{\text{max}}<\infty$; we will discuss the continuum limit below.

\subsection{Nearly-QP initial data}

We first study the nonlinear dynamics of initial data that closely
approximate a stable QP solution. We show that the simulation 
closely matches the linearized analysis, and we identify the origin of
deviations from linear behavior.

In anticipation of the following subsection, we define (a particular case of) 
two-mode initial data,
\begin{equation}\label{eq:2modeinit}
  E^{\text{two-mode}}_{j} = \frac{E}{2} (\delta_{j0} + \delta_{j1}),
\end{equation}
with the energy evenly divided between the two lowest modes. This data
has temperature $T=3.75$, and for later comparison with spacetime
simulations we take $E=0.0162$. There is therefore only one associated
QP solution, with $j_r=0$ (see Fig.~\ref{fig:SizeFamilies}). (We
neglect QP solutions that skip over modes.)
Following~\cite{Buchel:2014xwa}, we consider initial data that
interpolates between the two-mode initial data and the associated QP
solution,
\begin{equation}\label{eq:interpolation}
  E_j = (1 - \lambda)E^{\text{QP}}_j + \lambda E^{\text{two-mode}}_{j}
\end{equation}
where $E_j^{\text{QP}}$ is the associated QP spectrum. For all
$\lambda$, this interpolation preserves $E$ and $N$.

\begin{figure}[tb]
  \centering
    \includegraphics{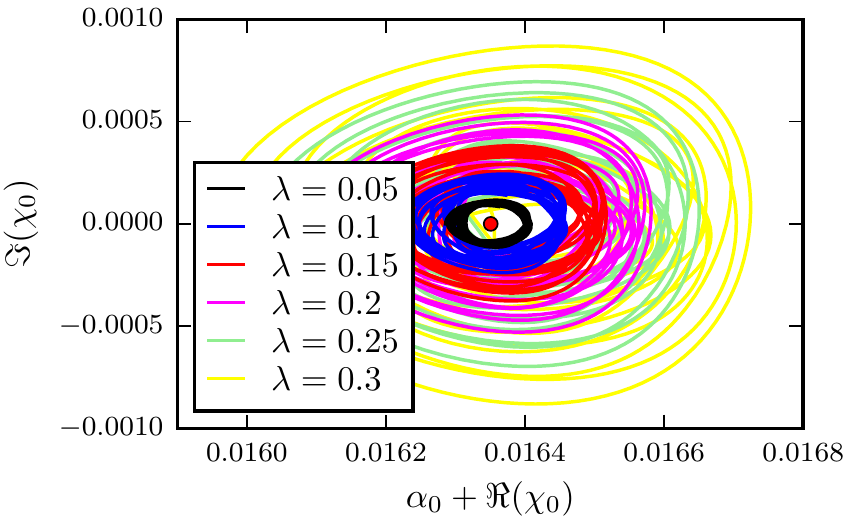}
    \caption{\label{fig:circleplotmode0}Evolution of mode $j=0$ in
      complex plane for interpolated initial data. The full solution
      here is
      $A_j(\tau)=[\alpha_j+\chi_j(\tau)]e^{-i(\beta_j+\tilde{\beta}_j^\lambda)\tau}$. Note
      that here (and subsequently) we also fit for a
      $\lambda$-dependent frequency shift satisfying
      $\tilde{\beta}_j^\lambda=\tilde{\beta}_0^\lambda+j(\tilde{\beta}_1^\lambda-\tilde{\beta}_0^\lambda)$,
      which arises from nonlinear effects (it is quadratic in
      $\lambda$). Had we not done so, there would be an additional
      overall phase oscillation. This phase, however, has no influence
      on the evolution of the energy spectrum, and tends to 0 as
      $\lambda\to 0$.}
  
\end{figure}

We performed nonlinear evolutions of the TTF equations~\eqref{eq:TTF}
for initial data~\eqref{eq:interpolation} as the parameter $\lambda$
was varied between $0.05$ and $0.30$.  Fig.~\ref{fig:circleplotmode0}
shows the oscillations of the lowest mode.
As $\lambda$ is increased, the mode continues to oscillate about the
QP solution, although with larger amplitude, as
expected.
\begin{figure}[tb]
  \centering
  \begin{tabular}{c}
    \subfloat{\label{fig:evolutionEnergyMode5Interpolation}\includegraphics{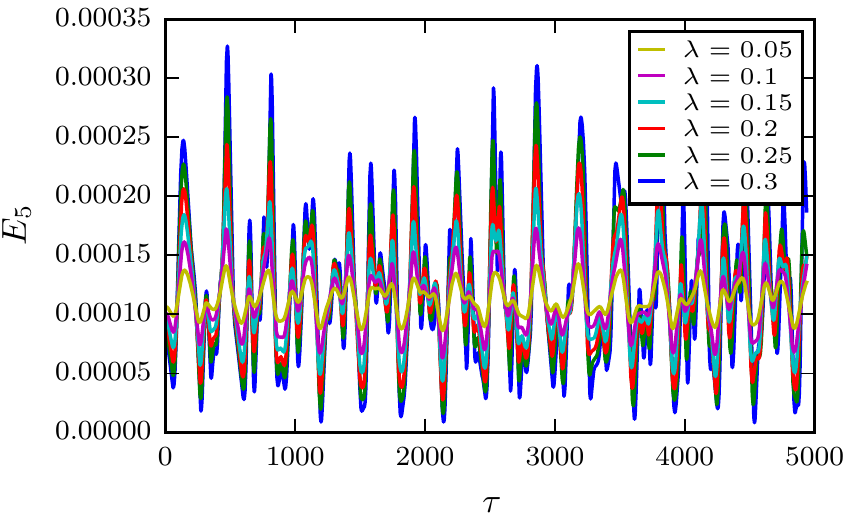}}\\
    \subfloat{\label{fig:FourierDecomposition}\includegraphics{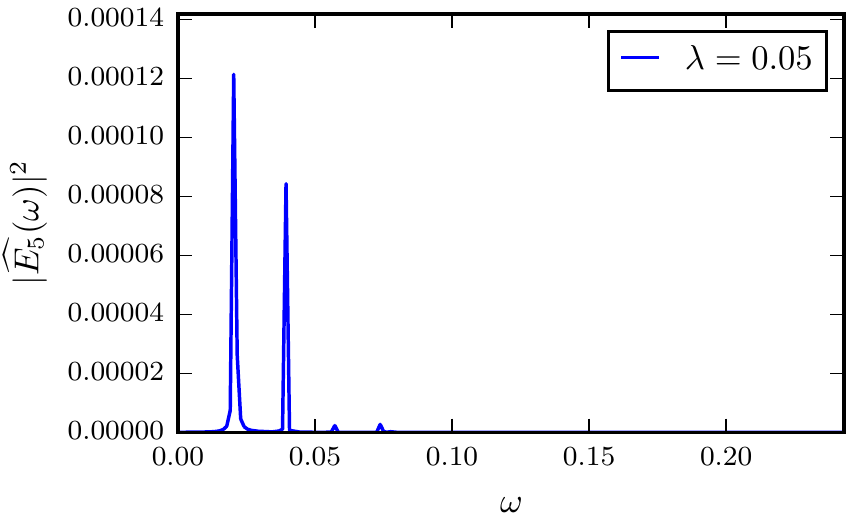}}
  \end{tabular}
  \caption{\label{fig:E5}Energy content of mode $j=5$ as a function of
    time (top), and its spectral density (bottom).}
\end{figure}
We plot the evolution of the energy of mode $j=5$ as $\lambda$ is
varied in Fig.~\ref{fig:evolutionEnergyMode5Interpolation}. This shows
that as the amplitude fluctuations increase considerably, the
periodicity is not significantly changed. The discrete Fourier
transform in Fig.~\ref{fig:FourierDecomposition} shows that the
oscillations are described by a discrete set of frequencies, as
expected from the linearized analysis.

\begin{figure}[tb]
  \centering
  \includegraphics{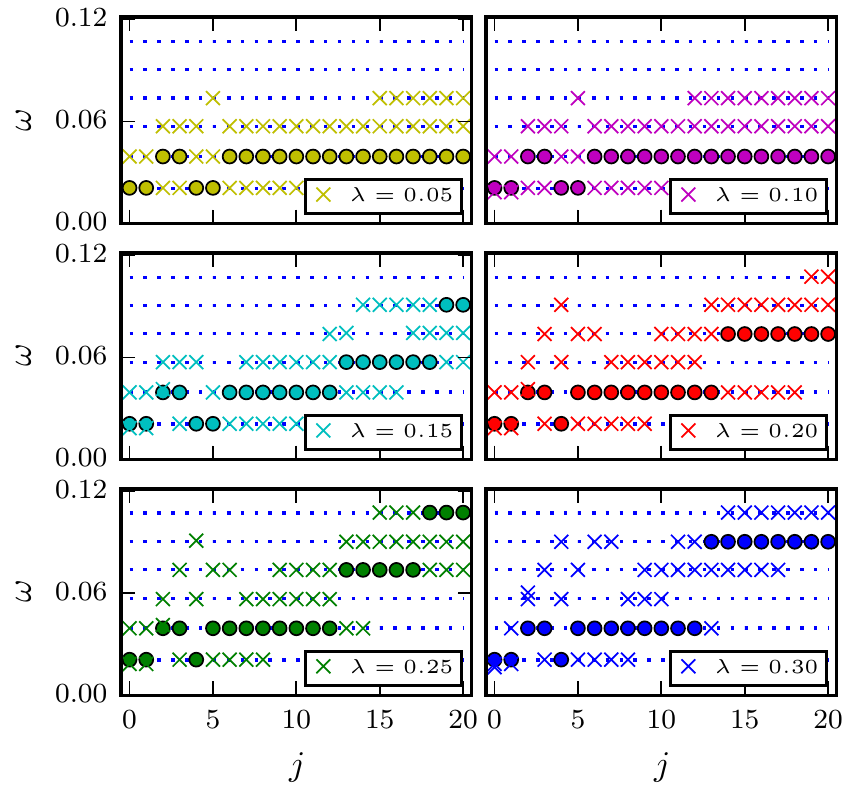}
  \caption{\label{fig:FrequenciesTable}Main oscillation frequencies
    present in the evolution of nearly-QP initial data (for various
    $\lambda$). Horizontal blue dotted lines represent the linear
    eigenfrequencies $\{|\lambda_n|\}$ of the QP solution. For each
    mode $j$, the three most dominant peaks of the spectral energy
    density are indicated by circles (largest peak) and crosses
    (secondary peaks). We also dropped secondary peaks if they were
    smaller than $1\%$ of the main peak. These plots were computed
    using a discrete Fourier transform on a regular grid of size
    $10^4$.}
\end{figure}
\begin{figure}[tb]
  \centering
  \includegraphics{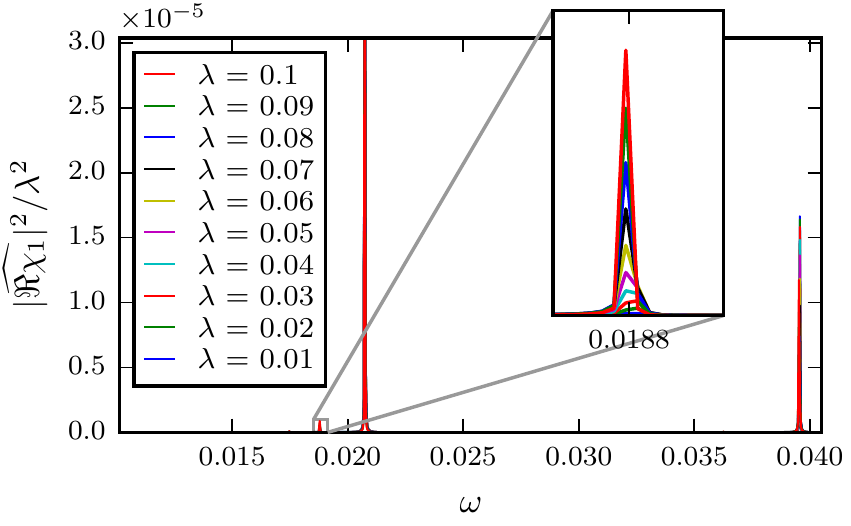}
  \caption{\label{fig:Frequencies-j0}Spectral energy density of
    $\Re(\chi_1)$, rescaled by $\lambda^2$. As $\lambda$ is varied,
    the heights of the two largest peaks are unchanged to leading
    order. Meanwhile, the rapid growth of the smallest peak (inset)
    with $\lambda$ indicates a nonlinear origin. Note also that
    $\lambda_1 - \lambda_0 = 0.01881$, which closely matches the
    position of the smallest peak. Thus, we conclude that the smallest
    peak arises from a quadratic coupling between modes $n=0,1$. Note
    that we also observed even smaller peaks at frequencies
    $\lambda_1 + \lambda_0$ and $\lambda_2 - \lambda_1$. These plots
    were computed using a discrete Fourier transform on a regular time
    grid with step size 0.25, up to time 150,000.}
\end{figure}
Fig.~\ref{fig:FrequenciesTable} shows the peaks of the spectral energy
density of $\Re(\chi_j)$ for $j\le20$. For the most part, these peaks
align closely with linear eigenfrequencies of the QP solution, but
there are several extraneous peaks at low frequencies. These arise
mostly in modes $j=0,1$ and for larger $\lambda$, which indicates they
arise nonlinearly. This is confirmed in Fig.~\ref{fig:Frequencies-j0},
which shows that the new peaks grow nonlinearly with $\lambda$. In
fact, the frequency of the first new peak is precisely the difference
between the two lowest eigenfrequencies $i\lambda_0=0.0207$ and
$i\lambda_1=0.0396$, so it is a nonlinear effect driven by a coupling
between the two lowest eigenmodes. More generally, given the
form~\eqref{eq:asympspectrum} of the spectrum,
\begin{equation*}
  i\lambda_n = C_1 + C_2n + O\left(\frac{1}{n}\right),
\end{equation*}
these lowest-frequency quadratically driven oscillations will arise at
frequencies that are approximately $C_2=0.0158$. At higher nonlinear
order, additional low-frequencies will appear, at, e.g.,
$C_2-C_1=0.0048$.

Notice also from Fig.~\ref{fig:FrequenciesTable} that larger-$j$ modes
are influenced more strongly by the larger-$n$ QP eigenmodes, as
expected from Fig.~\ref{fig:eigenmodesComponents}. Moreover, as the
deviation from the QP solution increases (larger $\lambda$) higher
frequency QP eigenmodes are excited.

This analysis shows that for nearly-QP initial data, the linearized
analysis of the associated QP equilibrium solution does an excellent
job of predicting the nonlinear dynamics, in particular the
periodicities. Furthermore, as $\lambda$ is increased further, an
additional low-frequency ($\approx C_2$) mode is nonlinearly excited
by the linear oscillations. This mode, we will see, is most closely
related to recurrences.

\subsection{Two-mode equal-energy initial data}\label{subsect:2Mode}

Setting $\lambda=1$ in the interpolated initial data of the previous
subsection, we obtain the two-mode equal-energy initial data, which
have received significant
attention~\cite{Balasubramanian:2014cja,Buchel:2014xwa}. As above,
$T=3.75$, so there is a single associated QP solution (that of the
previous subsection).

\begin{figure}[tb]
  \centering
  \includegraphics{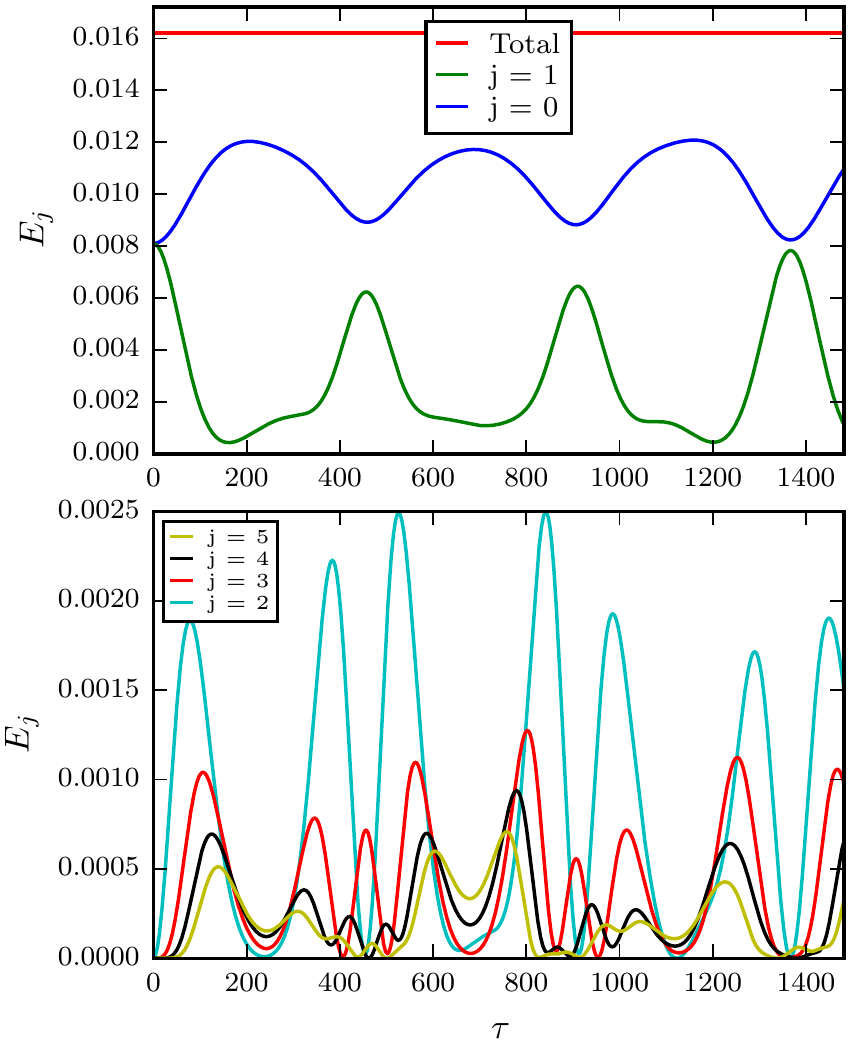}
  \caption{\label{fig:2modeEE-evolution}Energy evolution of first 6
    modes for two-mode equal-energy data with $T = 3.75$. The
    evolution was computed using the TTF equations with
    $j_{\text{max}} = 100$.}
\end{figure}
Fig.~\ref{fig:2modeEE-evolution} shows the nonlinear evolution of the
energy of the first six modes. The main recurrence time is closely
related to the periods of the QP eigenmodes, but is in fact slightly
longer. Indeed, the dominant time scale of the $j=1$ mode is
approximately $450$, and the first three linearized eigenmodes about
the QP solution have periods $303$, $159$ and $110$. In precisely the
manner described in the previous subsection for smaller $\lambda$,
nonlinear couplings between the eigenmodes drive oscillations at the
new (slightly longer, as compared to the largest eigenperiod, $303$)
characteristic time scale $2\pi/C_2\approx398$ (for
$j_{\text{max}}=100$). Notice also that at the third recurrence
($\tau\approx1350$) there is an even closer return to the initial
configuration, and that this coincides with a third order nonlinear
interaction time scale, $2\pi/(C_2-C_1)\approx1310$.

\begin{figure}[tb]
  \centering
    \includegraphics{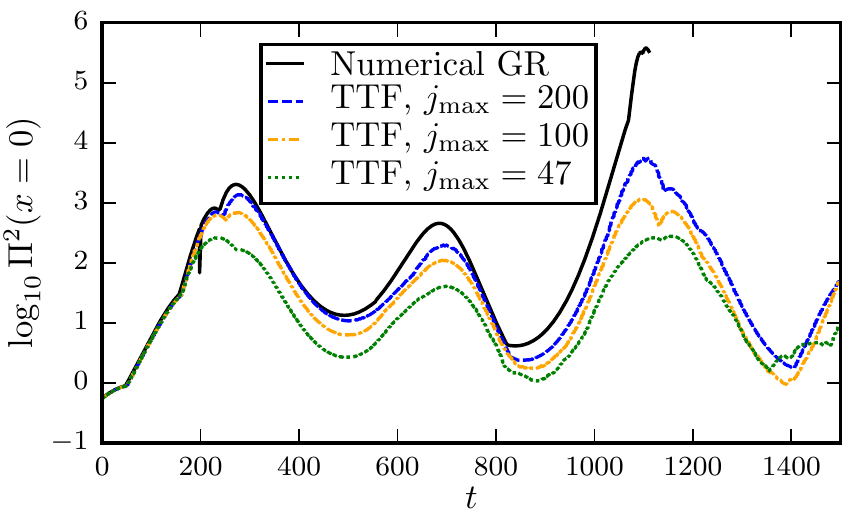}
    \caption{\label{fig:mopi-2modeEE}Upper envelope of $\Pi^2(x=0)$
      for two-mode initial data. Approach to full numerical relativity
      simulation is seen as $j_{\text{max}}$ is increased. The higher
      peaks arise because increasing $j_{\text{max}}$ allows the
      direct cascade to proceed to higher-$j$ modes, which are more
      peaked about $x=0$. This figure updates a similar figure in our
      previous work~\cite{Balasubramanian:2014cja}, with a higher
      resolution GR simulation, and also larger $j_{\text{max}}$ TTF
      simulations.}
\end{figure}
To compare with numerical simulations in AdS, we reconstruct the
spacetime fields from the TTF variables $\{A_j(\tau)\}$ (with
$\epsilon=1$ by convention to keep time axes consistent). We plot the
upper envelope of $\Pi^2 \equiv (\partial_t\phi)^2$ at the origin,
$x=0$, in Fig.~\ref{fig:mopi-2modeEE}. ($\Pi^2$ itself exhibits a
fast-time oscillation that is not of interest.) This quantity is
related to the Ricci scalar, and is frequently employed as an
indicator of collapse
(e.g.~\cite{Bizon:2011gg,Buchel:2012uh,Deppe:2014oua}). Notice that
$\Pi^2(x=0)$ can reach very large values in the course of evolution,
but inherits the recurrences from the energy plot. Growth in
$\Pi^2(x=0)$ reflects direct turbulent cascades of energy to high-$j$
modes, while decay reflects inverse cascades. The time scale of these
recurrences---troughs at $t=450,850$, peaks at $t=300,700,1100$---is
consistent with the predicted period of $2\pi/C_2\approx398$.

It is now clear that previously unexplained recurrence times can be
understood naturally as oscillations about QP equilibria, and they can
be predicted without any time-integrations. (In the case of the
two-mode data, the frequency\footnote{In contrast to the frequency,
  Fig.~\ref{fig:mopi-2modeEE} shows that the {\em amplitude} of
  recurrences depends strongly upon $j_{\text{max}}$ for the range we
  studied. Since the oscillation is nonlinear, there is no obvious way
  to predict this amplitude.} of recurrences emerges nonlinearly as
the asymptotic separation $C_2$ between eigenfrequencies.) Such
predictions are of particular relevance for their holographic
implications for field theories.

\subsection{Gaussian initial data, $\sigma = 4/10$}\label{subsubsec:Gaussian0p4}

Initial data with a Gaussian distribution for the scalar field in
position space have been closely scrutinized within the context of the
AdS stability problem (see,
e.g.,~\cite{Bizon:2011gg,Buchel:2013uba,Deppe:2014oua}). In
particular, collapse was first studied for a Gaussian with variance
$\sigma=1/16$. As noted in~\cite{Buchel:2012uh}, there is also a range
of $\sigma$ for which collapse is apparently averted. Armed with our
new understanding of perturbations about QP solutions, we here analyze
a non-collapsing Gaussian, and in the following subsection we study
the collapsing case.

\begin{figure}[tb]
  \centering
  \includegraphics{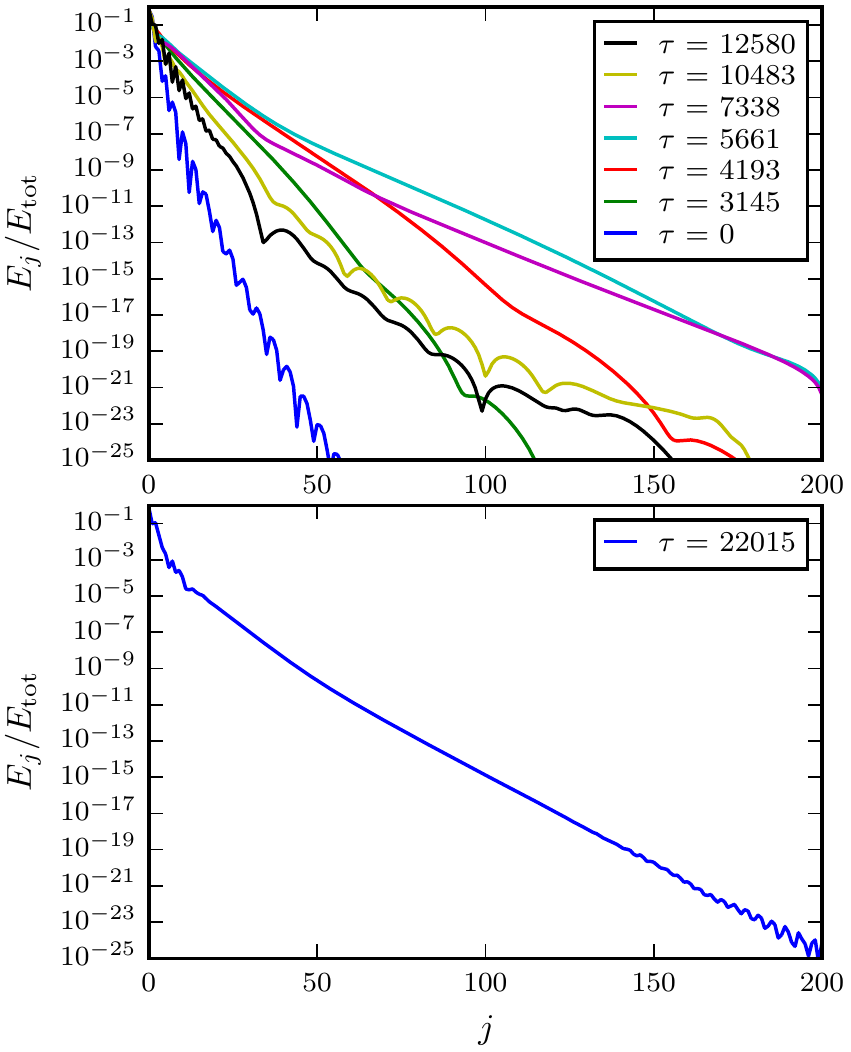}
  \caption{\label{fig:gaussian0p4energies}Evolution of the energy
    spectrum for $\sigma = 0.4$ Gaussian initial data. We show several
    times during the first direct and inverse cascades (top), and a
    much later time during the second inverse cascade
    (bottom). Spectra are all roughly exponential. ($j_{\text{max}} =
    200$)}
\end{figure}

The $\sigma=0.4$ Gaussian, which has $T=3.42$, is in many ways similar
to the two-mode initial data. There is a single associated QP
solution, and the evolution is characterized by a series of direct and
inverse cascades. Throughout the evolution, the energy spectrum (see
Fig.~\ref{fig:gaussian0p4energies}) remains roughly exponential---as
opposed to power-law---corresponding to non-collapse.

\begin{figure}[tb]
  \centering
  \includegraphics{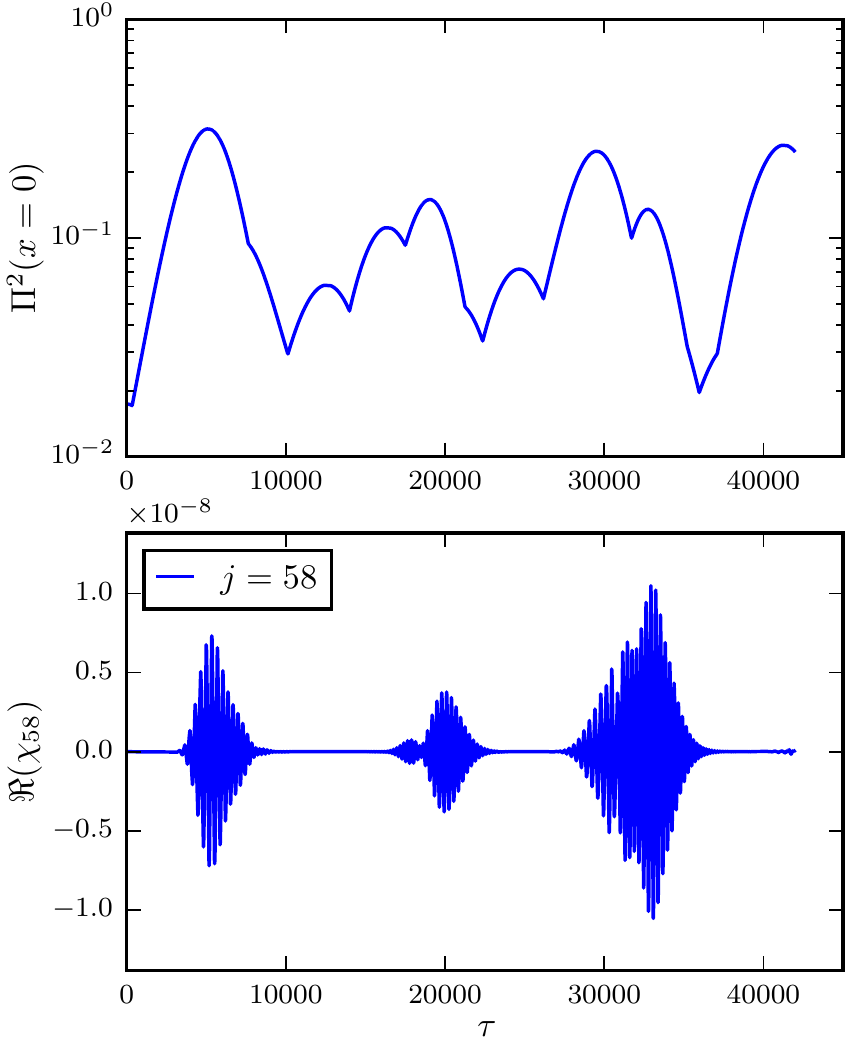}
  \caption{\label{fig:gaussian0p4realpartmode58}The evolution of
    $\Re(\chi_{58})$, and the upper envelope of $\Pi^2(x=0)$, for
    $\sigma=0.4$ Gaussian initial data. The high-frequency oscillation
    of $\Re(\chi_{58})$ is very well predicted by the linear analysis
    of the associated QP solution. The lower-frequency modulation is a
    combination of beating of linear modes and nonlinear driving, and
    it corresponds to the recurrences seen in
    $\Pi^2(x=0)$. ($j_{\text{max}} = 200$)}
\end{figure}
Observed oscillation periods can be predicted by analyzing the
associated QP solution in the same way as for the two-mode data. As a
representative example, we monitor the behavior of a high-frequency
($j = 58$) mode in Fig.~\ref{fig:gaussian0p4realpartmode58}. The
high-frequency oscillation of $\Re(\chi_{58})$ occurs with period
359, which to this accuracy matches exactly the period of one of the
$\lambda_n$. Similar agreement with the linearized frequencies can be
seen for the other modes.

The Gaussian data, however, differs from the two-mode initial data in
that, initially, it more strongly excites high-$j$ modes. In turn,
this causes increased excitation of high-$n$ QP eigenmodes. This is
reflected in complicated linear ``beating'' and nonlinear ``driving''
dynamics between excited modes seen in
Fig.~\ref{fig:gaussian0p4realpartmode58}. Here, the slow envelope
modulation arises as the difference in frequencies between subsequent
QP eigenmodes---with a corresponding period $2\pi/C_2$ for this QP
solution.  The amplitude of the beating is predicted by the linear
analysis to be $\sim 10\%$ of the measured amplitude and we expect
that nonlinear driving accounts for the remainder. Again, the
characteristic time scale is $2\pi/C_2$, which can be determined from
the linear spectrum to be $12500$ for $j_{\mathrm{max}} = 100$, in
agreement with Fig.~\ref{fig:gaussian0p4realpartmode58}. This time
scale also matches the recurrences in
Fig.~\ref{fig:gaussian0p4energies}.

\subsection{Gaussian initial data, $\sigma = 1/16$}

In contrast to all previous examples, the $\sigma=1/16$ Gaussian is
seen to collapse in numerical
simulations~\cite{Bizon:2011gg,Buchel:2012uh}. The temperature
$T=13.1$ suggests that there could in principle be several associated
QP equilibria, but nevertheless the data do not display any
oscillations, indicating that they are far from these equilibria.

\begin{figure}[tb]
  \centering
  \includegraphics{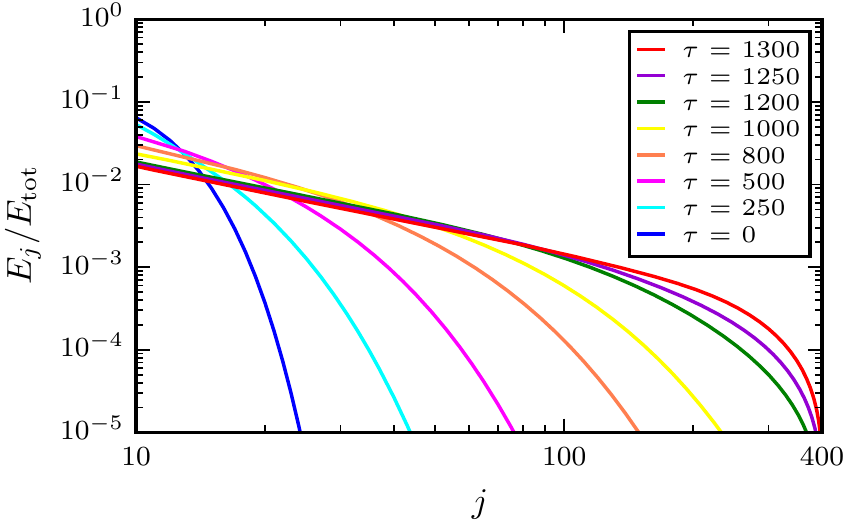}
  \caption{\label{fig:gaussian1o16varioustimes}Energy spectra at
    various times for $\sigma=1/16$ Gaussian initial data.  This
    figure shows the approach to power-law during the time evolution
    of the TTF equations. We used $j_{\text{max}} = 400$. Compare to the
    exponential spectra for the $\sigma=0.4$ Gaussian, illustrated in
    Fig.~\ref{fig:gaussian0p4energies}.}
\end{figure}
Consistent with previous full GR
simulations~\cite{Maliborski:2013via}, the energy spectrum of this
data in TTF approaches a power
law\footnote{Speculation~\cite{Basu:2015efa} that power laws do not
  arise in gravity (in an analogy to a self-interacting scalar field)
  are based on scaling assumptions for the
  $\mathcal{S}$-coefficients. In fact, the $\mathcal{S}$-coefficients
  grow with increasing mode number in gravity (see
  footnote~\ref{fn:scaling}), while they decay for the
  scalar~\cite{Basu:2015efa}, so the coupling to high modes is much
  stronger in gravity. This arises because of the spacetime
  derivatives present in the gravitational interaction.}
$E_j\sim(j+1)^{-\alpha}$ as it evolves in time (see
Fig.~\ref{fig:gaussian1o16varioustimes}). Extrapolating to
$j_{\text{max}}\to\infty$, such a spectrum would lead to diverging
spacetime fields (such as $\Pi^2$ at the origin), indicating the break
down of TTF as a valid description. At this point, higher order
dynamics have been seen to lead to
collapse~\cite{Bizon:2011gg,Buchel:2012uh}.  Despite the failure of
TTF to provide a valid description past the power law, the TTF
solution (for finite $j_{\text{max}}$) is perfectly well-defined for
longer times (beyond those shown in
Fig.~\ref{fig:gaussian1o16varioustimes}).

\begin{figure}[tb]
  \centering
  \includegraphics{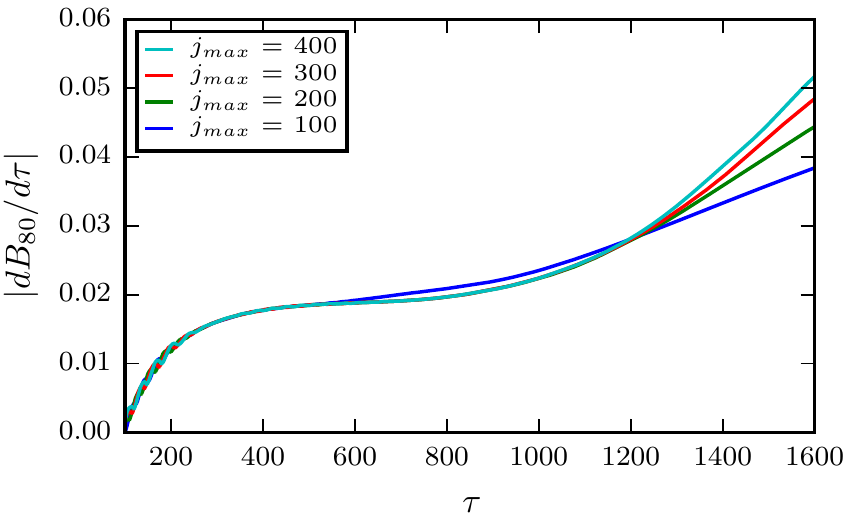}
  \caption{\label{fig:gaussian1o16phasederivative}Absolute value
    of the derivative of the phase of mode $80$, for an initial
    Gaussian $\sigma = \frac{1}{16}$ data, and different values of
    $j_{\text{max}}$.  We denote $A_j = R_j e^{i B_j (\tau)}$.}
\end{figure}
A recent publication~\cite{Bizon:2015} examined the TTF evolution of
two-mode data in AdS${}_5$, which displays a similar evolution to a
power-law spectrum as seen here. It was argued that in the limit
$j_{\text{max}}\to\infty$, TTF itself breaks down after the power law
is reached, with the time derivatives of phases of the mode amplitudes
$\propto\log(\tau-\tau_\ast)$. The truncated equations nevertheless
have a well-defined solution beyond $\tau_\ast$, which was described
as an unphysical ``afterlife.''  The authors of~\cite{Bizon:2015}
emphasized that even at finite $j_{\text{max}}$ a highly oscillatory
behavior led to numerical difficulties beyond $\tau_\ast$.  We note
that we did not encounter any numerical difficulties in our
simulations of $\sigma=1/16$ Gaussian data beyond this
time\footnote{\label{fn:scaling}The closed-form expressions of
  App.~\ref{sec:closedform} give rise to analytic expressions for the
  asymptotic scaling of the $\mathcal{S}$-coefficients. Precisely, in
  AdS${}_4$,
  $\mathcal{S}^{(j)}_{iji} \sim -\frac{128}{\pi} j^2 i^2 \ln{j}$. For
  comparison, Ref.~\cite{Bizon:2015} reports the corresponding
  expression in AdS${}_5$ to be $j^2 i^3$. Because of the $\ln j$
  factor, the arguments of~\cite{Bizon:2015} do not apply in
  AdS${}_4$; the phases (in their notation) $(B_n) \sim n\ln{n}$ can
  have $B_l + B_n - B_j - B_k \to \infty$ even for resonant quartets,
  so the ansatz taken for high modes is invalidated (e.g.,
  $B_{2i} + B_0 - 2B_{i} \sim i \ln{2}$). Since it would be natural
  for a logarithmic factor to arise in AdS${}_5$ as well, it would be
  useful to analytically compute in this case the asymptotic form of
  the $\mathcal{S}$-coefficients.} (see
App.~\ref{sec:numerical_schemes}).

We present in Fig.~\ref{fig:gaussian1o16phasederivative} the evolution
of the first derivative of the phase of mode $80$ for different values
of $j_\text{max}$. While the curves agree at early times, we were not
able to conclude whether they approach a limit near the collapse time
$\tau_\ast \approx 1500$ as $j_{\text{max}}\to\infty$. Nevertheless,
we do not observe the logarithmic blowup in the derivatives of the
phases reported in Fig.~5 of~\cite{Bizon:2015} for the case of
AdS${}_5$; in the present case of AdS${}_4$ the behavior seems less
extreme.

\section{Conclusions}

In this paper, we have analyzed perturbations of AdS$_4$ within the
TTF formalism. We identified a collection of two-parameter families of
QP equilibrium solutions to the TTF equations, and we established
their linear stability (in the sense of Lyapunov). For each QP
solution, this analysis gave rise to a new spectrum of eigenmodes
$\{\hat{\boldsymbol{e}}_n\}$---which are collective oscillations of
the AdS$_4$ normal modes $\{e_j\}$ about the QP solution---along with
their own oscillation frequencies $\{\lambda_n\}$. We also showed,
through several examples, that the linear analysis often remains valid
well into the nonlinear regime, and moreover, the leading nonlinear
effect is generally to introduce new frequencies that are combinations
of the $\{\lambda_n\}$.

A key takeaway message is that for initial data that do not collapse
as $\epsilon\to0$ in AdS$_4$ (or, at least, do not do so immediately),
{\em recurrences are simply oscillations about stable QP
  equilibria}. The relevant frequencies arise from the
$\{\lambda_n\}$. With our stability analysis, we now have a method of
predicting recurrence times without any need for time integrations.

Initial data that do collapse as $\epsilon\to0$ (in the sense that the
TTF description breaks down) are not sufficiently close to any QP
solution. We observed, in agreement with previous fully nonlinear
general relativistic simulations~\cite{Maliborski:2013via}, that these
data tend to approach power-law energy spectra. The presence of stable
QP equilibria thus reconciles the apparent tension between the fully
commensurate frequency spectrum of AdS and non-thermalizing initial
data. Indeed, fully commensurate frequency spectra would be expected
to thermalize as a result of the KAM theory and Arnold
diffusion~\cite{lichtenberg1992regular}, yet the QP solution [in
conjunction with conservation of $(E,N,H)$] constrains the available
phase space and acts as an island of stability. The spectra of
perturbations about QP solutions are non-resonant themselves (except
asymptotically for large $n$), indicating that thermalization should
not be expected within a stable island.

We point out two facts that play a role in extending this work to
systems beyond AdS$_4$. In AdS, each QP family is parametrized by the
two conserved quantities $E$ and $N$. In contrast, for other confining
systems, such as a flat spherical cavity~\cite{Maliborski:2012gx},
additional resonances are present, and $N$ is not conserved. As a
result, in that case we expect only {\em one}-parameter families of
stable equilibria. Meanwhile, as the dimensionality of the system is
increased, couplings to high-$j$ modes become stronger
($\mathcal{S}$-coefficients become larger), which then drive stronger
turbulent cascades. These effects compete in deciding collapse versus
non-collapse.

\acknowledgements We would like to thank O. Evnin for discussions and
comments on the manuscript, and A.~Buchel for discussions and comments
throughout this project.  This work was supported by the NSF under
grant PHY-1308621~(LIU), by NASA under grant NNX13AH01G, by NSERC
through a Discovery Grant (to L.L.) and by CIFAR (to L.L.).  Research
at Perimeter Institute is supported through Industry Canada and by the
Province of Ontario through the Ministry of Research and Innovation.
A.M. thanks the Perimeter Institute for Theoretical Physics for
hospitality and accommodations, as part of the Graduate Fellows
Program, during an internship sponsored by the \'Ecole Normale
Sup\'erieure.

\appendix

\section{Closed form for the $\mathcal{S}$-coefficients}\label{sec:closedform}

Reference~\cite{Craps:2014jwa} provided new simplified formulas for
the $\mathcal{S}$-coefficients described by integrals of products of
the mode functions~\eqref{eq:ej}. Notice that our conventions for the
$\mathcal{S}$-coefficients differ from those in~\cite{Craps:2014jwa}
by a factor $4$, that is to get our coefficients
$\mathcal{S}_{i j k l}$ one has to multiply the expressions given in
\cite{Craps:2014jwa} by $4$. Note also that in this section, for
commodity, we rewrite
$\mathcal{S}_{i j k l} \equiv \mathcal{S}^{(i)}_{j k l}$.

Here we shall give new closed-form formulas found for the tensors of
\cite{Craps:2014jwa}, which allowed us to compute the $\mathcal{S}$-coefficients
up to $j_{\text{max}} = 400$ in a short time.

Recall the expressions for $\mathcal{S}_{i j k l}$ given
in~\cite{Craps:2014jwa}:
\begin{align}
\label{eq:S-diagonal}
\mathcal{S}_{l l l l} =  2& \omega_l^2 X_{l l l l} + 6 Y_{l l l l} + 8\omega_l^4 W_{l l l l} + 8\omega_l^2 W^{*}_{l l l l} \nonumber \\&- 4 \omega_l^2 (A_{l l} + \omega_l^2 V_{l l}),
\end{align} 
and, for $i \neq l$,
\begin{align}
\label{eq:S-twoequal}
\mathcal{S}_{l i l i} &= \mathcal{S}_{l i i l} = 2(\frac{\omega_i^2+\omega_l^2}{\omega_l^2-\omega_i^2})(\omega_l^2 X_{i l l i} - \omega_i^2 X_{l i i l}) \nonumber \\&+ 8(\frac{\omega_l^2 Y_{i l i l} - \omega_i^2 Y_{l i l i}}{\omega_l^2-\omega_i^2}) + 4(\frac{\omega_i^2\omega_l^2}{\omega_l^2-\omega_i^2})(X_{i l l i} - X_{l i l i}) \nonumber \\& + 2(Y_{i i l l} + Y_{l l i i}) + 4\omega_i^2\omega_l^2(W_{l l i i} + W_{i i l l}) + \omega_i^2 W^{*}_{l l i i} \nonumber \\& + \omega_l^2 W^{*}_{i i l l} - \omega_l^2 (A_{i i} + \omega_i^2 V_{i i}) \, .
\end{align} 
and finally, if $i \neq l$ and $i \neq k$,
\begin{align}
\label{eq:S-generalcase}
&\mathcal{S}_{i j k l} =\nonumber
\\& -(\frac{1}{\omega_i + \omega_j} + \frac{1}{\omega_i - \omega_k} + \frac{1}{\omega_j - \omega_k})(\omega_i\omega_j\omega_k X_{l i j k} - \omega_l Y_{i l j k}) \nonumber 
\\&-(\frac{1}{\omega_i + \omega_j} + \frac{1}{\omega_i - \omega_k} - \frac{1}{\omega_j - \omega_k})(\omega_j\omega_k\omega_l X_{i j k l} - \omega_i Y_{j i k l}) \nonumber 
\\&-(\frac{1}{\omega_i + \omega_j} - \frac{1}{\omega_i - \omega_k} + \frac{1}{\omega_j - \omega_k})(\omega_i\omega_k\omega_l X_{j i k l} - \omega_j Y_{i j k l}) \nonumber 
\\&-(\frac{1}{\omega_i + \omega_j} - \frac{1}{\omega_i - \omega_k} - \frac{1}{\omega_j - \omega_k})(\omega_i\omega_j\omega_l X_{k i j l} - \omega_k Y_{i k j l}).
\end{align} 
The quantities that appear in these coefficients are defined by integrals of the mode functions (recall that we work here in  $d = 3$ spatial dimensions):
\begin{align}
\label{eq:XIntegral}
X_{i j k l} = \int_{0}^{\frac{\pi}{2}} \! e'_i(x) e_j(x) e_k(x) e_l(x) \frac{\sin^3{x}}{\cos{x}}\, \mathrm{d}x ,
\end{align} 
\begin{align}
\label{eq:YIntegral}
Y_{i j k l} = \int_{0}^{\frac{\pi}{2}} \! e'_i(x) e_j(x) e'_k(x) e'_l(x) \frac{\sin^3{x}}{\cos{x}}\, \mathrm{d}x  ,
\end{align} 
\begin{align}
\label{eq:WIntegral}
W_{i i l l} = \int_{0}^{\frac{\pi}{2}} \mathrm{d}x \, e_i(x)^2\sin{x}\cos{x} \int_0^x  \mathrm{d}y \, e_k(y)^2 \sin{y}\cos{y}  ,
\end{align} 
\begin{align}
\label{eq:WStarIntegral}
W^*_{i i l l} = \int_{0}^{\frac{\pi}{2}} \mathrm{d}x \, e'_i(x)^2\sin{x}\cos{x} \int_0^x  \mathrm{d}y \, e_k(y)^2 \sin{y}\cos{y}  ,
\end{align} 
\begin{align}
\label{eq:VIntegral}
V_{i j} = \int_{0}^{\frac{\pi}{2}} \mathrm{d}x \, e_i(x) e_j(x) \sin{x}\cos{x}  ,
\end{align} 
\begin{align}
\label{eq:AIntegral}
A_{i j} = \int_{0}^{\frac{\pi}{2}} \mathrm{d}x \, e'_i(x) e'_j(x) \sin{x}\cos{x} .
\end{align} 

To simplify these expressions, we used the form of the mode functions
$e_j(x)$. We know that~\eqref{eq:ej}
\begin{equation}
 e_j(x) = 4\sqrt{\frac{(j+1)(j+2)}{\pi}}\cos^3x\,{}_2F_1\left(-j,3+j;\frac{3}{2};\sin^2x\right). \nonumber
\end{equation}
Since the first argument of the hypergeometric function is a negative
integer, the hypergeometric function appearing in $e_j$ is in fact a
polynomial function of degree $j$. Using then the expansion for the
hypergeometric function,
\begin{align}\label{eq:modenotsimplified}
{}_2F_1\left(-j,3+j;\frac{3}{2};x\right) = \sum\limits_{k=0}^{j} \binom{j}{k} (-1)^k \frac{(3+j)_k}{(\frac{3}{2})_k} x^k,
\end{align}  
with $(a)_k$ the rising Pochhammer symbol, the following identity holds as proven below: 
\begin{align}\label{eq:modesimplified}
e_j(x) = \frac{1}{\sqrt{\pi (j+1)(j+2)}} \frac{f_{j+1}(x)}{\sin(x)};
\end{align}  
with 
\begin{align}\label{eq:fm}
f_m(x) = (m+1)\sin(2 m x) + m\sin[2 (m+1) x].
\end{align}
To establish this result, first notice that in $d = 3$, the mode
functions satisfy the differential equation,
\begin{align}\label{eq:diffeqmode}
e''_i(x) + 2\left[\tan(x)+\cot(x)\right]e'_i(x) + (2 i+3)^2 e_i(x) =0 .
\end{align}
Next, using~\eqref{eq:modesimplified} and~\eqref{eq:ej}, it is
immediate to check that at $x=0$ the two expressions and their
derivatives have the same limits,
\begin{align}\label{eq:limitmodes}
&\lim_{x\to 0} e_i(x) = 4 \sqrt{\frac{(i+1)(i+2)}{\pi}} \nonumber ,
\\&\lim_{x\to 0} e'_i(x) = 0 \nonumber ,
\\&\lim_{x\to 0} e''_i(x) = - \frac{4 \omega_i^2}{3} \sqrt{\frac{(i+1)(i+2)}{\pi}} .
\end{align}
It is then straightforward to check that
both~\eqref{eq:modesimplified} and~\eqref{eq:modenotsimplified}
satisfy the differential equation~\eqref{eq:diffeqmode} on
$\left(0,\frac{\pi}{2}\right)$. At this point, while one might be
tempted to conclude both expressions are the same, we note
that~\eqref{eq:diffeqmode} is singular at $x=0$ and
$x=\frac{\pi}{2}$. We thus proceed as follows: let us first denote
$e_j(x) \equiv \cos(x) u_j(\sin^2x)$. Next, with the expanded form of
both~\eqref{eq:modenotsimplified} and~\eqref{eq:modesimplified} [using
$\sin(2 m x) = \sum_{k = 0}^{m-1} \binom{2m}{2k+1} (-1)^k
\sin(x)^{2k+1} \cos(x)^{2m-2k-1}$]
we can show that $u_j(t)$ is in both cases a polynomial function of
$t$; the remaining task is to show both polynomials are the same. To
check this fact, denote by $\{Q(t),T(t)\}$ the polynomial equal to
$u(t)$ from expressions~\eqref{eq:modenotsimplified}
and~\eqref{eq:modesimplified}, respectively. We can then substitute
each (multiplied by $\cos x$) into Eq.~\eqref{eq:diffeqmode}. The
resulting equation for both $Q$ and $T$ in $\left(0,1\right)$ is the
same simple differential equation,
\begin{align}\label{eq:diffeqpolynom}
&4t(1-t)^2 f''(t) + 2(t-1)(4t-3)f'(t) \nonumber
\\&  + (\omega_i^2 (1-t) -3 +t) f(t) =0 ,
\end{align}
with $f$ standing for either $Q$ or $T$. Now, since $Q$ and $T$ are
polynomials, they verify this equation everywhere.  Moreover, this
equation gives rise to the following order-2 relation on the
coefficients of $Q$ and $T$, denoting $Q(X) = \sum\limits_k q_k X^k$
and $T(X) = \sum\limits_k t_k X^k$, which is
\begin{align}\label{eq:recurrencerelation}
&2(k+1)(2k+3)u_{k+1} + (\omega_j^2 - 3 -8k^2 -6k)u_k \nonumber
\\&+ (4k(k-1) + 1 - \omega_j^2)u_{k-1} = 0 .
\end{align}

Consequently, according to~\eqref{eq:recurrencerelation}, $(q_k)_k$
and $(t_k)_k$ are both uniquely determined by the same relation, and
by their first two values. It is thus sufficient to check that
$q_0 = t_0$ and $q_1 = t_1$, which is given by~\eqref{eq:limitmodes}
[since $e(0) = u(0)$ and $e''(0) = -u(0) + 2u'(0)$]. We have thus
proven that $Q = T$, that is that~\eqref{eq:modesimplified} is a valid
expression for $e_j$.

Let us also stress that these calculations, done in AdS${}_4$, are not
straightforwardly extended to other dimensions. Some inspection and
analysis of the calculations in different dimensions, however, points
towards a similar simplification of eigenmodes in odd spatial
dimensions $d$, though we have not exhaustively studied this question.

We then have, for the derivative, 
\begin{align}\label{eq:derivativemodesimplified}
e'_j(x) = \frac{3+2 j}{\sqrt{\pi (j+1)(j+2)}}\frac{\cos(x)}{\sin^2(x)} g_{j+1}(x) ,
\end{align}  
with 
\begin{align}\label{eq:gm}
g_m(x) = -(m+1)\sin(2 m x) + m\sin[2 (m+1) x] .
\end{align}

Since the indefinite integrals appearing in $W$ and $W^*$ are easy to
compute, one can now reduce the problem to computing many integrals of
the type
$\int_0^{\frac{\pi}{2}} \mathrm{d}x \, x^\gamma \cos^\alpha{x}
\sin^\beta{x} \times F(2 m x)$
and
$\int_0^{\frac{\pi}{2}} \mathrm{d}x \, x^\gamma \cos^\alpha{x}
\sin^\beta{x} \times F((2 m +1) x)$,
with $F$ the cosine or the sine, m an integer, $\gamma \in \{0,1\}$, and
$\alpha$ and $\beta$ integers greater or equal to $-1$.

We give here some conventions and the few delicate integrals that one
has to compute in order to get the relevant coefficients. We will
denote $\delta_m$ the Kronecker delta function, and $\text{Sign}(m)$ the
function taking the value $1$ on $\mathbb{N}^*$, the value $-1$ on
$\mathbb{Z}_-^*$, and $\text{Sign}(0) = 0$. Some of these integrals were
found thanks to several formulas found in~\cite{Gradshteyn}.

We also denote $\psi$ the polygamma function
$\psi(x) = \frac{\Gamma'(x)}{\Gamma(x)}$ where $\Gamma$ is the Euler
function. We denote $p \equiv m+n$ and $k \equiv m-n$:
\begin{align*}
\forall n \in \mathbb{Z} &, \\ &\int_0^{\frac{\pi}{2}} \mathrm{d}x \, \frac{\sin{((2n+1)x)}}{\sin{(x)}} = \frac{\pi}{2}(\text{Sign}(n) + \delta(n))  ,
\end{align*}
\begin{align*}
&\forall m \in \mathbb{N} , \\ &\int_0^{\frac{\pi}{2}} \mathrm{d}x \, x \frac{\cos{((2 m + 1) x)}}{\sin{(x)}} = \frac{\pi}{4} (-1)^m\left[\psi(\frac{m+2}{2}) - \psi(\frac{1+m}{2})\right] ,
\end{align*}
\begin{align*}
 \forall (m,n) & \in \mathbb{N}^2 , \\ &\int_0^{\frac{\pi}{2}} \mathrm{d}x \, \frac{\cos{(x)}}{\sin{(x)}} \sin{(2 m x)} \sin(2 n  x) = 
\\& \begin{cases}
		\frac{1}{2} \left[\psi(\frac{1+p}{2}) - \psi(\frac{1+k}{2})\right] & \text{if p is even} \\
		-\frac{1}{2} \left[-\psi(\frac{p}{2}) + \psi(\frac{k}{2}) + \frac{1}{k} - \frac{1}{p}\right]       & \text{if p is odd}
		\end{cases}
\end{align*}
We are then able to find closed form expressions for every quantity
we need. Nevertheless, these expressions appear to be too long for the
$X$ and $Y$ tensor to be written clearly on one page, and are
therefore not given here. We shall now give the expressions found for
$A$, $V$, $W$ and $W^*$.
\begin{widetext}

  Due to the symmetric property of $A$ and $V$ it is sufficient to
  restrict to $i \geq j$. For the case where $i+j$ is an even integer,
\begin{align}
&A_{i j} = \frac{(2 i+3) (2 j+3)}{2 \pi  \sqrt{(i+1) (i+2) (j+1) (j+2)}}\left((2 i (i+3)+2 j (j+3)+7)\left[\psi(\frac{i+j+3}{2}) - \psi(\frac{i-j+1}{2})\right]\right.\nonumber
\\&\left.-\frac{2 (i+1) (j+1) \left(-2 (i-4) j^2-2 i (i+8) j+i (2 i (i+4)-3)+2 j^3-3 j-13\right)}{\left((i-j)^2-1\right) (i+j+3)}\right) ,
\end{align}
\begin{align}
V_{i j} = \frac{1}{2 \pi  \sqrt{(i+1) (i+2) (j+1) (j+2)}} &\left((2 i+3) (2 j+3)\left[\psi(\frac{i+j+1}{2})-\psi(\frac{i-j+1}{2})\right]-\frac{(2 i+3)^2}{i+j+3}+\frac{i (i+2)}{i-j-1}+ \right.\nonumber
\\& \left.\frac{(i+1) (i+3)}{-i+j-1}+\frac{6-8 i (i+1)}{i+j+1}+4 (3 i+4)\right) .
\end{align}
\bigskip
For the case where $i+j$ is an odd integer, and $i \geq j$,
\bigskip
\begin{align}
A_{i j} = \frac{(2 i+3) (2 j+3)}{2 \pi  \sqrt{(i+1) (i+2) (j+1) (j+2)}} &\left((2 i (i+3)+2 j (j+3)+7)(\psi(\frac{i+j+2}{2}) - \psi(\frac{i-j}{2}))-4 i j-\frac{4 i (i+3)+7}{i-j} \right.\nonumber
\\&\left.+\frac{7 i (i+2)+6}{i+j+2}+\frac{(i+1) (i+3)}{i+j+4}-8 i\right) ,
\end{align}
\begin{align}
&V_{i j} = \frac{1}{4 \pi  (i-j) \sqrt{(i+1) (i+2) (j+1) (j+2)} (i+j) (i+j+2) (i+j+4)}\left(-8 (5 i+7) j^4-4 (i (14 i+85)+93) j^3+\right.\nonumber
\\&\left.8 (i ((i-21) i-93)-89) j^2+4 (i (i (i (6 i+37)+19)-102)-108) j \right.\nonumber
\\&\left.-2 (2 i+3) (2 j+3) (i-j) (i+j) (i+j+2) (i+j+4) \left[\psi(\frac{i-j}{2})-\psi(\frac{i+j}{2})\right]+16 i (i+1) (2 i (i+5)+9)\right) .
\end{align}
Last, we have the following general values for $W$ and $W^*$:
\begin{align}
&W_{m m n n} =\frac{(-2 m^2 (2 l+3)+4 m (l^2-2)+2 l (3 l+5)+3)}{16 \pi  (m+1) (m+2) (l+1) (l+2) (2 l+3)} \delta_{m-l}
-\frac{(2 m+3)^2 (2 l (l+3)+5)}{16 \pi  (m+1) (m+2) (l+1) (l+2) (2 l+3)} \text{Sign}(m-l) \nonumber
\\&+ \frac{(2 m+3)^2}{4 \pi  (m+1) (m+2)}(-\psi(m+1)+\psi(m+\frac{3}{2})+2 \ln (2))\nonumber
\\&-\frac{1}{16 \pi  (m+1)^2 (m+2)^2 (l+1) (l+2) (2 l+3)}\left[8 m^4 (l+1) (2 l (l+4)+7)+8 m^3 (l+1) (l (14 l+55)+48)\right.\nonumber
\\&\left.+m^2 (4 l (l (73 l+355)+527)+979)+m (4 l (17 l (5 l+24)+602)+1113)+2 (l (l (74 l+351)+515)+237)\right] ,
\end{align}
\begin{align}
&W^*_{m m n n} =-\frac{(2 m+3)^2}{16 \pi  (m+1) (m+2) (n+1) (n+2) (2 n+3)}\left[4 m^4 (2 n+3)-8 m^3 ((n-3) n-7)-2 m^2 (2 n (6 n (n+6)+41)-3)\right.\nonumber
\\&\left. +4 m (n (n (6 n (n+3)-13)-54)-18)+3 (2 n (n (6 n^2+28 n+41)+21)+9)\right]\delta_{m-n} \nonumber
\\&-\frac{(m+2) (2 m+3)^2 (n+1) (-m+n+1)^2}{4 \pi  (m+1) (n+2) (2 n+3)} \delta _{m-n-1} + \frac{(m+1) (2 m+3)^2 (n+2) (m-n+1)^2}{\pi  (m+2) (n+1) (8 n+12)} \delta _{m-n+1} \nonumber
\\&-\frac{(2 m+3)^2 \left(4 m^2 (2 n (n+3)+5)+12 m (2 n (n+3)+5)-2 n (n+3) (8 n (n+3)+27)-37\right)}{16 \pi  (m+1) (m+2) (n+1) (n+2) (2 n+3)} \text{Sign}(m-n) \nonumber
\\&-\frac{(2 m+3)}{16 \pi  (m+1)^2 (m+2)^2 (n+1) (n+2) (2 n+3)}\left[16 m^5 (n+1) (2 n (n+4)+7)+8 m^4 (n+1) (2 n (17 n+67)+117) \right.\nonumber
\\&\left.+m^3 (8 n (n (n (93-4 n)+517)+797)+3018)+m^2 (4 n (n (n (187-36 n)+1458)+2408)+4701)\right. \nonumber
\\&\left.+m (4 n (n (n (31-52 n)+936)+1736)+3545)+2 (n (n (423-2 n (24 n+31))+947)+513)\right] .
\end{align}
\end{widetext}

\section{Methods for obtaining quasi-periodic solutions}\label{sec:QPspace}

Here we describe the two approaches we took to generate the families
of QP solutions in Sec.~\ref{sec:qp-solutions-newton}.

\subsection{Direct solution using Newton-Raphson method}

Given an appropriate starting point, the Newton-Raphson method
provides successively better approximate solutions to a set of coupled
equations. Thus, to find numerical QP solutions of~\eqref{eq:QP} it is
necessary to choose an appropriate ``seed'' for the algorithm.

Although we parametrized QP solutions by $E$ and $N$ in the main text,
it is more appropriate here to choose parameters from among $\beta_0$,
$\beta_1$ and $\{\alpha_j\}$. For example, to find $j_r=0$ solutions
we fix $\alpha_0=1$ (an arbitrary choice because of the scaling
symmetry) and $\alpha_1\ll\alpha_0$. Eqs.~\eqref{eq:QP} may be solved
to eliminate $\beta_0$ and $\beta_1$, leaving $j_{\text{max}}-1$
equations and $j_{\text{max}}-1$ unknowns. For the remaining
variables, we choose an exponential energy spectrum as a seed,
\begin{equation}
  \alpha_j \sim \frac{3e^{-\mu j}}{2j+3}, \label{expanzats}
\end{equation}
with $\mu=\log{[{3}/{(5\alpha_1)}]}$. For sufficiently small
$\alpha_1$ the Newton-Raphson method gives back a solution with nearly
exponential energy spectrum, as seen in Fig.~\ref{fig:qp-jr0}.

As $\alpha_1$ is increased, the QP energy spectra deform away from
exponentials and it becomes increasingly difficult to find solutions
with the exponential ansatz (\ref{expanzats}). In fact,
in~\cite{Balasubramanian:2014cja} we could not find solutions with
$\alpha_1>0.42$. Slightly better results can be obtained by taking
$(\beta_1-\beta_0)/\beta_0$ as a parameter ($\beta_1\gg\beta_0$
approaches the single-mode solution), however, this also breaks down
for large $T$. To fully uncover the $j_r=0$ family we require the
technique of the following subsection.

Solutions within $j_r>0$ families can be obtained in a similar manner,
now fixing $\alpha_{j_r}=1$ and $\alpha_{j_r+1}\ll \alpha_{j_r}$. To
pick a seed for the Newton-Raphson algorithm, we solve the first
several QP equations~\eqref{eq:QP} perturbatively in
$\alpha_{j_r+1}/\alpha_{j_r}$.

\subsection{Perturbation from known solution}

Now suppose $A_j^{\text{QP}}(\tau)=\alpha_je^{-i\beta_j\tau}$ is a
known numerical QP solution. Sec.~\ref{sec:nearbyQP} shows that there
is in general a 2-parameter family of perturbations to nearby QP
solutions, so by following such perturbations new QP
solutions---otherwise not readily obtainable through the Newton-Raphson
method---can be constructed.

Since one of the parameters is, as usual, an overall scale, there is
only one nontrivial parameter. It is, therefore, convenient to fix $N$
and vary $E$ by a small amount $\delta E$. Following
Sec.~\ref{sec:nearbyQP}, we numerically solve the linear system of
equations,
\begin{eqnarray}
  \label{eq:QPgen1}0&=&2\omega_j\left[\alpha_j(\theta_1+\omega_j\theta_2)+\beta_j u_j\right]\\
  &&+\sum_{klm}\mathcal{S}^{(j)}_{klm}\left(\alpha_l\alpha_m u_k+\alpha_k\alpha_m u_l+\alpha_k\alpha_l u_m\right),\nonumber\\
  \label{eq:QPgen2}0&=&8\sum_j\omega_j\alpha_j u_j,\\
  \label{eq:QPgen3}\delta E&=& 8\sum_j\omega_j^2\alpha_j u_j,
\end{eqnarray}
for the variables $(\theta_1,\theta_2,\{ u_j\})$. We then update the QP solution,
\begin{eqnarray}
  \alpha_j&\to&\alpha_j+ u_j,\\
  \beta_j&\to&\beta_j+\theta_1+\omega_j\theta_2.
\end{eqnarray}
The new QP solution has particle number $N$, energy $E+\delta E$, and
therefore the temperature has changed by $\delta T = \delta E/ N$.

With the updated QP solution, the procedure may be iterated repeatedly
to obtain finite-sized $\Delta T$. (The Newton-Raphson method can be
used periodically to ensure the deviation from actual QP solutions
does not become too large.) In this manner, we obtained the full QP
families illustrated in Fig.~\ref{fig:SizeFamilies}. These families
terminate when solutions to \eqref{eq:QPgen1}--\eqref{eq:QPgen3} no
longer exist (i.e., when the associated matrix has vanishing
determinant).

\section{Minimization of $H$ and linear stability of QP
  solutions}\label{sec:ddH}
We shall here show the relation between the minimization of $H$ for a
QP solution and its linear stability. We know from~\eqref{eq:HMinimal}
that $H$ has a critical point at a QP solution. Here we compute the
second order change in $H$.

Let us take a generic second order perturbation of a QP solution, that
does not perturb $E$ and $N$,
\begin{equation}\label{eq:Order2Perturbation}
\alpha_j \rightarrow \alpha_j + A^{(1)}_j + A^{(2)}_j ,
\end{equation}
where $A^{(k)}_j$ is the order $k$ perturbation of $\alpha_j$.
Recall the expression~\eqref{eq:H} of $H$,
\begin{equation}
 H \equiv -\frac{1}{4}\sum_{jklm}\mathcal{S}^{(j)}_{klm}\bar{A}_j\bar{A}_k A_l A_m - \frac{E}{4}\sum_{j}\mathcal{C}_j|A_j|^2 . \nonumber
\end{equation} 
Inserting~\eqref{eq:Order2Perturbation} into this equation, one finds
\begin{align}
 &-\delta^2 H = \frac{E}{4}\sum\limits_j \mathcal{C}_j \left[\alpha_j ( A^{(2)}_j + \bar{A}^{(2)}_j) + |A^{(1)}_j|^2 \right] \nonumber 
\\&+\frac{1}{4} \sum\limits_{j,k,l,m} \mathcal{S}^{(j)}_{klm} \left[\alpha_j \alpha_k \alpha_l A^{(2)}_m + \alpha_j \alpha_k \alpha_m A^{(2)}_l + \alpha_j \alpha_l \alpha_m \bar{A}^{(2)}_k \right.\nonumber
\\&\left. + \alpha_k \alpha_l \alpha_m \bar{A}^{(2)}_j + \alpha_l \alpha_m \bar{A}^{(1)}_j \bar{A}^{(1)}_k + \alpha_j \alpha_k A^{(1)}_l A^{(1)}_m \right.\nonumber
\\&\left.+ \alpha_k \alpha_m \bar{A}^{(1)}_j A^{(1)}_l +\alpha_k \alpha_l \bar{A}^{(1)}_j A^{(1)}_m +\alpha_j \alpha_m \bar{A}^{(1)}_k A^{(1)}_l \right.\nonumber
\\&\left.+\alpha_j \alpha_l \bar{A}^{(1)}_k A^{(1)}_m \right] \nonumber \, .
\end{align} 
Let us concentrate on the part where only $A^{(2)}_j$ appears. Using
the QP TTF equation~\eqref{eq:QP}, as well as the
relations~\eqref{eq:Sspliting} and~\eqref{eq:Rspliting} on the
$\mathcal{S}$ coefficients, one can reduce the expression of this part
to
\begin{equation} \label{eq:partdHA2}
\sum\limits_k \mathcal{C}_k \alpha_k^2 \sum\limits_j \omega_j^2 \alpha_j (\bar{A}^{(2)}_j + A^{(2)}_j) - \sum\limits_j \omega_j \alpha_j \beta_j (\bar{A}^{(2)}_j + A^{(2)}_j)  .
\end{equation}
Now, since $E$ and $N$ are conserved at both linear and quadratic
level, we have, for the quadratic level,
\begin{align}
  &\sum\limits_j \omega_j^2 \left[\alpha_j ( A^{(2)}_j + \bar{A}^{(2)}_j) + |A^{(1)}_j|^2\right] = 0 \nonumber  ,
  \\& \sum\limits_j \omega_j \left[\alpha_j ( A^{(2)}_j + \bar{A}^{(2)}_j) + |A^{(1)}_j|^2\right] = 0\nonumber  ,
\end{align} 
which can be rewritten as 
\begin{align}
&\forall (u_j) \in \mathbb{R}^\mathbb{N} \text{ s.t. } u_j = u_0 + j (u_1 - u_0), \nonumber
\\& \sum\limits_j \omega_j u_j \left[\alpha_j ( A^{(2)}_j + \bar{A}^{(2)}_j) + |A^{(1)}_j|^2\right] = 0 \nonumber \, .
\end{align} 
Using this identity in~\eqref{eq:partdHA2}, one can rewrite the full
second order variation of the Hamiltonian as a function of the linear
perturbation only,
\begin{align}
 &-\delta^2 H = \frac{E}{4}\sum\limits_j \mathcal{C}_j |A^{(1)}_j|^2 + \sum\limits_j \omega_j \beta_j |A^{(1)}_j|^2 \nonumber 
\\&- \sum\limits_k \mathcal{C}_k \alpha_k^2 \sum\limits_j \omega_j^2 \alpha_j |A^{(1)}_j|^2 \nonumber
 \\&+\frac{1}{4} \sum\limits_{jklm} \mathcal{S}^{(j)}_{klm} \left[\alpha_l \alpha_m \bar{A}^{(1)}_j \bar{A}^{(1)}_k + \alpha_j \alpha_k A^{(1)}_l A^{(1)}_m \right.\nonumber
\\&\left.+ \alpha_k \alpha_m \bar{A}^{(1)}_j A^{(1)}_l +\alpha_k \alpha_l \bar{A}^{(1)}_j A^{(1)}_m +\alpha_j \alpha_m \bar{A}^{(1)}_k A^{(1)}_l \right.\nonumber
\\&\left.+\alpha_j \alpha_l \bar{A}^{(1)}_k A^{(1)}_m \right] \nonumber
\end{align}
Now, with~\eqref{eq:Sspliting}, one can reduce this last expression to 
\begin{align}
 &-\delta^2 H = \frac{E}{4}\sum\limits_j \mathcal{C}_j |A^{(1)}_j|^2 + \sum\limits_j \omega_j \beta_j |A^{(1)}_j|^2 \nonumber 
\\&- \sum\limits_k \mathcal{C}_k \alpha_k^2 \sum\limits_j \omega_j^2 \alpha_j |A^{(1)}_j|^2  +\sum\limits_{jklm} \mathcal{S}^{\text{S}}_{jklm} \alpha_k \alpha_l \bar{A}^{(1)}_j A^{(1)}_m  \nonumber
 \\&+\frac{1}{4} \sum\limits_{jklm} \mathcal{S}^{\text{S}}_{jklm} \alpha_l \alpha_m (\bar{A}^{(1)}_j \bar{A}^{(1)}_k +  A^{(1)}_j A^{(1)}_k)\nonumber .
\end{align}
Let us now rewrite the linear perturbation $A^{(1)}_j$ in terms of real and imaginary part,
\begin{equation}
\delta A_j = R_j + i I_j \nonumber  .
\end{equation}
If we denote by $X$ the column vector
$(R_0,\dots,R_{j_{max}},I_0,\dots,I_{j_{max}})$, and $M$ the matrix
such that we have $-\delta^2 H = X^{T} M X$, them $M$ is of the simple
form $\begin{pmatrix*}[r]
  A' & 0 \\
  0 & B'
 \end{pmatrix*}$, where $A'$ and $B'$ are both square matrices of size $j_{\text{max}} +1$, and we have the following expressions for their coefficients: 
\begin{align}
A'_{i,j} &= \frac{1}{2} \sum\limits_{lm} \mathcal{S}^{\text{S}}_{ijlm} \alpha_l \alpha_m + \omega_j \beta_j \delta_{i,j} + \frac{E}{4}\mathcal{C}_j \delta_{i,j} \nonumber
\\&- \omega_j^2 \delta_{i,j} \sum\limits_k \mathcal{C}_k \alpha_k^2 + \sum\limits_{kl} \mathcal{S}^{\text{S}}_{iklj} \alpha_k \alpha_l  ,
\\B'_{i,j} &= -\frac{1}{2} \sum\limits_{lm} \mathcal{S}^{\text{S}}_{ijlm} \alpha_l \alpha_m + \omega_j \beta_j \delta_{i,j} + \frac{E}{4}\mathcal{C}_j \delta_{i,j} \nonumber
\\&- \omega_j^2 \delta_{i,j} \sum\limits_k \mathcal{C}_k \alpha_k^2 + \sum\limits_{kl} \mathcal{S}^{\text{S}}_{iklj} \alpha_k \alpha_l \, .
\end{align}
Note that these matrix elements are quite similar to the matrix
elements of the matrix $A$ whose elements can be deduced
from~\eqref{eq:gamma} and~\eqref{eq:delta}.

Indeed, writing $A$ in the form $\begin{pmatrix*}[r]
  0 & -C \\
  D & 0
 \end{pmatrix*}$, one can, with the same type of calculations, prove the following simple identities:
\begin{align}
&\label{eq:Aprime}A'_{i,j} = \omega_i D_{i,j} - 2 \alpha_i \alpha_j (\omega_j^2 \mathcal{C}_i -\omega_i^2 \mathcal{C}_j) , \\
&\label{eq:Bprime}B'_{i,j} = \omega_i C_{i,j} .
\end{align}
In~\eqref{eq:Aprime}, since we are interested in the sign of
$X^{T} A' X$ to characterize stability, the right antisymmetric part
will play no role and we can ignore it. Let us also recall that we are
interested in the sign of $X^{T} M X$, with $X$ satisfying the linear
conservation of $E$ and $N$, that is, if
$X = (R_0,\dots,R_{j_{\text{max}}},I_0,\dots,I_{j_{\text{max}}})$ ,
\begin{align}\label{eq:linearconservation}
&\forall (u_j) \in \mathbb{R}^\mathbb{N} \text{ s.t } u_j = u_0 + j (u_1 - u_0), \nonumber
\\& \sum\limits_j \alpha_j \omega_j u_j R_j = 0 \nonumber ,
\end{align} 
which is equivalent to saying that $(R_j)$ is orthogonal to the
vectors $x_1 \equiv (\alpha_j \omega_j)$ and
$x_2 \equiv (\alpha_j \omega_j^2)$ in the Euclidean
$\mathbb{R}^{j_{max} +1 }$ space. We will thus place ourselves in the
two spaces $E \equiv (x_1,x_2)^{\perp}$ for $A'$ and $D$, and
$F \equiv \mathbb{R}^{j_{\text{max}} +1 }$ for $B'$ and $C$.

We note that in order to get the announced result, one has to assume
that $C$ and $D$ are both diagonalizable.  We have seen numerically
this is the case, but we have not rigorously proven this.

Let us now assume that $H$ has a local minimum at the QP solution
$(\alpha_j)$. Then that means that $A'$ and $B'$ are negative,
\begin{align}
& \forall X \in E, X^{T} A' X \leq 0  \, ,
\\& \forall Y \in F, Y^{T} B' Y \leq 0 \, .
\end{align}
Denoting $T \equiv \begin{pmatrix*}[r]
  \omega_0 & 0 & 0 \\
	0 & \dots & 0 \\
  0 & 0 & \omega_{j_{\text{max}}}
 \end{pmatrix*}$, we have $A' = T D$. Taking any eigenvector $X$ of $D$ with eigenvalue $\lambda$ (since $D$ is diagonalizable), we have $X^{T} A' X = \lambda \sum\limits_i \omega_i X_i^2  \leq 0$, which means that $Sp(D) \subset \mathbb{R}_{-}$. Using the exact same trick one can show that, 
\begin{equation}\label{th:symnegative}
T D \text{ is a negative symmetric matrix} \Leftrightarrow Sp(D) \subset \mathbb{R}_{-} .
\end{equation}

But, by computing the expression for the $D$ coefficients, it is
immediate that $D T^{-1}$ is also symmetric.  Now, since we know that
$A' = T D$ is a negative matrix, we deduce that $D T^{-1}$ is also
negative. Since $T C$ is also negative, and since the non-zero
eigenvalues of the product of two negative matrices are positive, the
real eigenvalues of $D C$ are all positive. Notice that since $D C$
and $C D$ have the same characteristic polynomial, this is also the
case for $C D$.

Let us recall that we argued that our system
\eqref{eq:gamma}--\eqref{eq:delta} is stable if and only if the
eigenvalues of $A$ are all pure imaginary. This means, since by
deriving \eqref{eq:gamma}--\eqref{eq:delta} again one can decouple
the system of equations, that the eigenvalues of $A^2$ are all real
and negative. But since $A^2 = \begin{pmatrix*}[r]
  -C D & 0 \\
  0 & -C D
\end{pmatrix*}$
, we know that the spectrum of $A^2$ is going to be in
$\mathbb{R}_{-}$ if $H$ has a minimum at the QP solution. So we know
that if $H$ has a minimum at a QP solution then this solution is
linearly stable.

Notice that this reasoning also holds if $H$ has a maximum at a QP
solution; in that case the solution will have only unstable modes (we
however never observed such a solution).

\section{Numerical integration method}\label{sec:numerical_schemes}
The integration of the TTF equations~\eqref{eq:TTF} requires special
care as, depending on the values of the coefficients $A_j$, they can
become stiff.  Stiff equations can be handled with explicit
methods---where the timestep must be small enough to ensure
stability---or implicit methods---where stability issues can be more
easily avoided but care must be exercised so as not to ``discard''
relevant short-time-scale physics by adopting too large a timestep.
We have implemented both explicit and implicit methods as well as
performed self-convergence in our analysis to ensure the correctness
of the obtained results.

\begin{figure}[tb]
  \centering
  \includegraphics{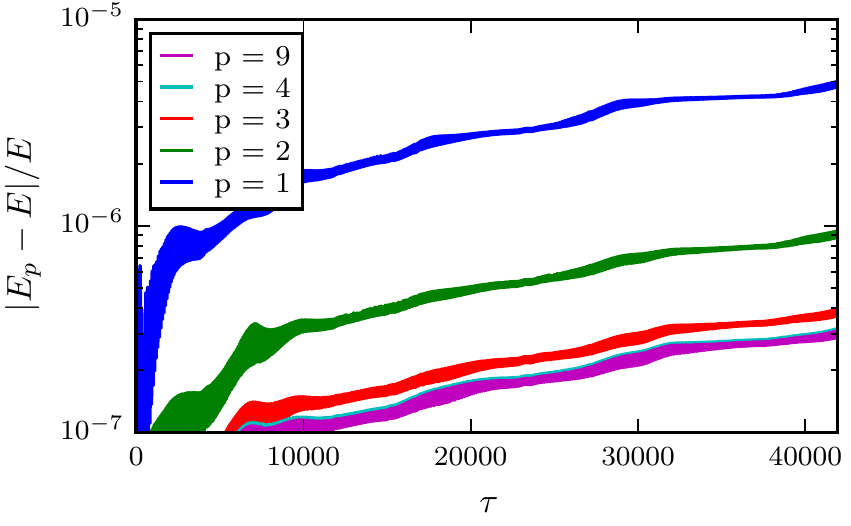}
  \caption{\label{fig:ErrorEnergy}The error in the total energy $E$
    for a Gaussian initial data of variance $\sigma = 0.4$, using
    different values of $p$. We used $j_{\text{max}} = 100$ for
    these calculations}
\end{figure}

In particular, we have employed the explicit (predictor-corrector)
Adams method as well as backward differentiation formulas (both with
adaptive timestepping) and, as in~\cite{Bizon:2015}, the implicit
Runge-Kutta scheme of order 6.  As an illustration, we present here
two tests of the validity of the implicit scheme adopted and our
strategy to ensure no relevant short-time-scale is discarded. We held
fixed the double-precision employed and varied the precision of our
adaptive step size method by $16p/10$ digits, $p=1,\ldots,10$.
Fig.~\ref{fig:ErrorEnergy} illustrates the change in conserved energy
$E$ vs integration time for different values of $p$. As is evident in
the figure, the error quickly converges to a limit function, which is
already reached for $p=4$. (The remaining error is due to the double
precision numbers.) We note that the results presented through the
paper have been obtained with $p=5$ with the implicit method.

\begin{figure}[tb]
  \centering
  \includegraphics{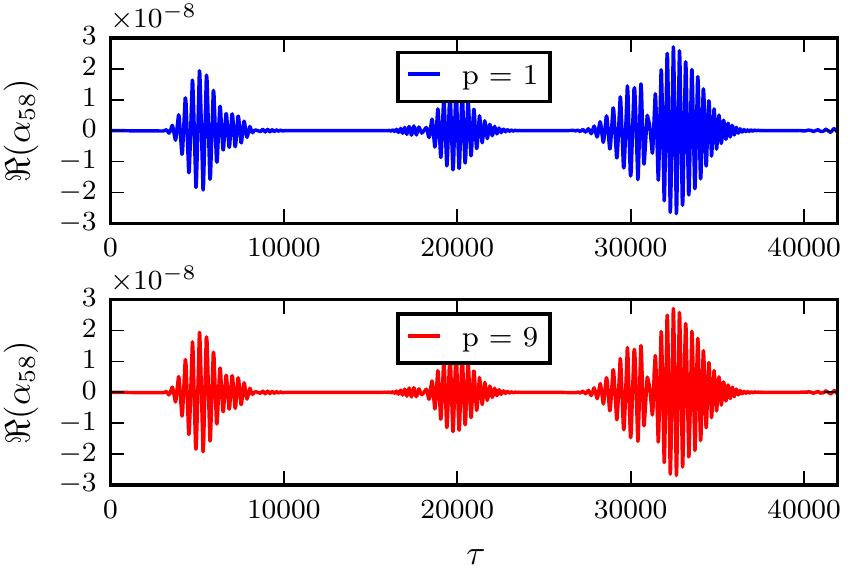}
  \caption{\label{fig:ErrorMode58}The evolution of the real part of
    the $50^{\text{th}}$ mode, for the extremal values of $p$ we
    took, 1 and 9. The difference between the two curves is of order
    $10^{-10}$. We used $j_{\text{max}} = 100$ for these
    calculations}
\end{figure}
To further illustrate that no relevant short-time-scale physics was
accidentally discarded by the use of an implicit integration scheme,
we show in Fig.~\ref{fig:ErrorMode58} the evolution of a
representative mode $j=50$ mode for two rather distinct values of
$p$. The difference between these two figures is of order $10^{-10}$.

\bibliography{references} 
\end{document}